\documentclass[prd,superscriptaddress]{revtex4}

\pdfoutput=1

\usepackage{amssymb}
\usepackage{graphicx}
\usepackage{dcolumn}
\usepackage{bm}
\usepackage{amssymb,amsmath,bm}  
\usepackage[usenames, dvipsnames]{color}
\usepackage{hyperref}
\usepackage{multirow}
\usepackage[utf8]{inputenc}
\usepackage{balance}
\usepackage{enumitem}
\usepackage{lipsum}
\usepackage{graphicx}
\usepackage{placeins}
\usepackage{xparse}
\usepackage{cancel}	
\usepackage{subfigure}
\usepackage[normalem]{ulem}

\newcommand{\perMpc}{{\rm Mpc^{-1}}}
\newcommand{\pr}{^{\prime}}
\newcommand{\pp}{^{\prime\prime}}
\newcommand{\Rcal}{\mathcal{R}}
\newcommand{\Dphi}{\Delta \phi}

\newcommand{\fint}{\int_{-\infty}^\infty df}
\newcommand{\Oint}{\int d^2\hat{\Omega}}
\newcommand{\bk}{\mathbf{k}}
\newcommand{\vO}{\hat{\Omega}}
\newcommand{\bx}{\mathbf{x}}

\newcommand{\dpsi}{\Delta \psi}
\newcommand{\Pcal}{\mathcal{P}}

\begin{document}

\title{Finding the chiral gravitational wave background of an axion-SU(2) inflationary model using CMB observations and laser interferometers}

\author{Ben Thorne}
\affiliation{%
  Kavli Institute for the Physics and Mathematics of the Universe (Kavli IPMU, WPI), UTIAS, The University of Tokyo, Chiba, 277-8583, Japan}
\affiliation{%
  Oxford Astrophysics, Denys Wilkinson Building, Keble Road, Oxford, OX1 3RH, United Kingdom}
%***
\author{Tomohiro Fujita}
\affiliation{%
  Stanford Institute for Theoretical Physics and Department of Physics, Stanford University, Stanford, CA 94306, USA}
\affiliation{%
  Department of Physics, Kyoto University, Kyoto, 606-8502, Japan}
%***
\author{Masashi Hazumi}
\affiliation{%
  Kavli Institute for the Physics and Mathematics of the Universe (Kavli IPMU, WPI), UTIAS, The University of Tokyo, Chiba, 277-8583, Japan}
\affiliation{%
  Institute of Particle and Nuclear Studies, KEK, 1-1 Oho, Tsukuba, Ibaraki 305-0801, Japan}
\affiliation{%
  SOKENDAI (The Graduate University for Advanced Studies), Hayama, Miura District, Kanagawa 240-0115, Japan}
\affiliation{%
  Institute of Space and Astronautical Studies (ISAS),
  Japan Aerospace Exploration Agency (JAXA), Sagamihara, Kanagawa 252-5210, Japan}
%***
\author{Nobuhiko Katayama}
\affiliation{%
  Kavli Institute for the Physics and Mathematics of the Universe (Kavli IPMU, WPI), UTIAS, The University of Tokyo, Chiba, 277-8583, Japan}
%***
\author{Eiichiro Komatsu}
\affiliation{%
  Kavli Institute for the Physics and Mathematics of the Universe (Kavli IPMU, WPI), UTIAS, The University of Tokyo, Chiba, 277-8583, Japan}
\affiliation{%
  Max Planck Institute for Astrophysics, Karl-Schwarzschild-Str. 1, 85748 Garching}
%***
\author{Maresuke Shiraishi}
\affiliation{%
  Kavli Institute for the Physics and Mathematics of the Universe (Kavli IPMU, WPI), UTIAS, The University of Tokyo, Chiba, 277-8583, Japan}
\affiliation{%
  Department of General Education, National Institute of Technology, Kagawa College, 355 Chokushi-cho, Takamatsu, Kagawa 761-8058, Japan}

\begin{abstract}
  A detection of B-mode polarization of the Cosmic Microwave Background (CMB) anisotropies would 
  confirm the presence of a primordial gravitational wave background (GWB).
  In the inflation paradigm this would be an unprecedented probe of the 
  energy scale of inflation as it is directly proportional to the power spectrum of the GWB. 
  However,  similar tensor perturbations can be produced by the matter fields 
  present during inflation, breaking the simple relationship between energy 
  scale and the tensor-to-scalar ratio $r$. It is therefore important to find 
  ways of distinguishing between the generation mechanisms of the GWB. 
  Without doing a full model selection, we analyse the 
  detectability of a new axion-SU(2) gauge field model by calculating the 
  signal-to-noise of future CMB and interferometer observations 
  sensitive to the chirality of the tensor spectrum. 
  We forecast the detectability of the resulting CMB temperature and B-mode 
  (TB) or E-mode and B-mode (EB) cross-correlation 
  by the LiteBIRD satellite, considering the effects of residual foregrounds, 
  gravitational lensing, and assess the ability of 
  such an experiment to jointly detect primordial TB and EB spectra and 
  self-calibrate its polarimeter. We find that LiteBIRD will be able to detect
  the chiral signal for $r_*>0.03$ with $r_*$ denoting the tensor-to-scalar ratio at the peak scale, and that the maximum signal-to-noise 
  for $r_*<0.07$ is $\sim 2$. 
  We go on to consider an advanced stage of a LISA-like mission, which is 
  designed to be sensitive to the intensity and polarization of the 
  GWB. We find that such experiments would complement
  CMB observations as they would be able to detect the chirality of the GWB
  with high significance on scales inaccessible to the CMB. 
  We conclude that CMB two-point statistics are limited in their ability to distinguish this 
  model from a conventional vacuum fluctuation model of GWB generation, due 
  to the fundamental limits on their sensitivity to parity-violation. 
  In order to test the predictions of such a model as compared to vacuum 
  fluctuations it will be necessary to test deviations from the self-consistency
  relation, or use higher order statistics to leverage the non-Gaussianity of the model. 
  On the other hand, in the case of a spectrum peaked at very small scales inaccessible to the CMB, a highly 
  significant detection could be made using space-based laser interferometers.
\end{abstract}

  \date{\today}
  \maketitle

\section{Introduction}
  \label{sec:intro}
  
  Over the past two decades the temperature and polarization
  anisotropies of the Cosmic Microwave Background (CMB) have 
  been measured with increasing sensitivity, ushering in the 
  era of `precision cosmology'. It is the aim of the next 
  generation of CMB experiments to better measure the 
  polarization of the CMB in order to 
  detect its primordial B-mode polarization, parametrized by $r$,
  the ratio between tensor and scalar perturbations, which would provide 
  strong evidence for the presence of a primordial gravitational 
  wave background (GWB) (see e.g.\ \cite{baumann:2009, Kamionkowski:2015yta, Guzzetti:2016mkm} for review).
  Normally, the GWB is produced only by quantum fluctuations 
  of the vacuum during inflation, 
  and is consequently simply related to the energy 
  density of inflation
  : $\rho_{\inf}^{1/4}\sim \left(\frac{r}{0.01}\right)^{1/4} 
  1.04 \times 10^{16} \ {\rm GeV}$. A measurement of the power spectrum of 
  tensor perturbations to the metric would therefore be an extremely 
  powerful probe of physics at GUT scales $\sim 10^{16} \ {\rm GeV}$.
  
  Given the importance of this measurement, many experiments are 
  currently making observations of the polarized CMB, such as 
  POLARBEAR \cite{polarbear:2014}, SPTPol \cite{sptpol:2015}, 
  ACTPol \cite{actpol:2014}, BICEP2 / Keck Array \cite{bkp16}, 
  and Planck \cite{planck_inflation:2016}.
  The best current observational constraints come from a
  combination of BICEP2/Keck and Planck (BKP) data to give $r < 0.07$ (95\% C.L) 
  \cite{bkp16}, but the next round of CMB experiments, such as the 
  LiteBIRD satellite \cite{Hazumi:2012aa}, the CORE satellite \cite{core:2015}
   and the ground-based 
  Stage-4 \cite{cmbs4} effort, seek to push constraints on $r$ to 
  $\sim 10^{-3}$.
  Interestingly, this search for B-modes may also be sensitive to 
  the dynamics of subdominant fields other than the inflaton, considering the 
  possibility of alternative gravitational wave generation scenarios.
  Some particular matter fields present during inflation can produce   primordial tensor perturbations similar to those
  sourced 
  by vacuum fluctuations. 
  Therefore, in the event of a detection 
  of $r$, we must first understand its source. 

  Recent efforts to provide alternative models for the generation 
  of gravitational waves, which are also consistent with existing observations,
  have introduced the coupled system of the 
  axion and gauge fields as the spectator sector in addition to the inflaton sector \cite{namba/etal:2016,dimastrogiovanni/etal:2016, obata/etal:2016, ferreira/etal:2016, caprini/sorbo:2014, mukohyama/etal:2014}. Such a setup is quite natural from the point of view of particle physics, since many high energy theories contain axion fields  and its coupling to some gauge fields, namely the Chern-Simons term: $(\chi/f)F^{\mu \nu}\tilde{F}_{\mu \nu}$. In particular, string theory typically predicts the existence of numerous axion fields. From the view point of low energy effective field theory, at the same time, such dimension five interaction term is expected to  exist, because
it respects the shift symmetry of the axion field, $\chi \rightarrow \chi + {\rm constant}$. Therefore it is strongly motivated to investigate the observational consequence of their dynamics during inflation in  light of the role of inflation as a unique probe of high energy physics.
  
  Interestingly enough, the GWB produced by the additional axion-gauge sector 
  has several characteristic features, including non-Gaussianity, scale-dependence, and chirality. 
  A model involving a U(1) gauge field was studied first, and it was confirmed that the resulting GWB is amplified to the same level as the scalar perturbation \cite{namba/etal:2016,peloso/etal:2016} and hence visible in CMB B-mode observations \cite{shiraishi/etal:2016} and interferometer experiments \cite{garcia-bellido/etal:2016}. Recently, a more intriguing model due to a SU(2) gauge field was also examined, achieving a surpassing GWB production against the scalar sector \cite{dimastrogiovanni/etal:2016}. This yields more rich phenomenology, and thus motivates us toward the assessment of its detectability.

%--- phenomenology due to chiral GW

  Gravitational waves may be decomposed into modes with left (L) and right (R) handed polarization. 
        A GWB produced by conventional vacuum fluctuations would have equal amplitudes of L and R, but the effect of the Chern-Simons term in the theory is to allow their amplitudes to differ
        \cite{Lue:1998mq,namba/etal:2016,dimastrogiovanni/etal:2016}. Such a chiral GWB
          would have signatures observable both in CMB polarization and by laser interferometers.
         CMB polarization may be decomposed into modes of opposing parity: E and B \cite{kamionkowski/etal:1997, zaldarriaga/seljak:1997}. A detection of a correlation between 
         E and B modes (EB), or between temperature and B modes (TB), would therefore be strong evidence of a parity-violating GWB \cite{Lue:1998mq,saito/etal:2007,glusevic/kamionkowski:2010,gerbino/etal:2016,shiraishi/etal:2016}. 
         To-date observational constraints using the CMB are consistent with no parity-violation and are dominated by systematic uncertainty 
         \cite{gerbino/etal:2016,planck_parity:2016, gruppuso/etal:2016, molinari/etal:2016}.
        An alternative to using the CMB is to directly probe the circular polarization of the GWB, denoted with the circular 
        polarization Stokes parameter $V(f)$, using gravitational 
        interferometers. Interferometers are sensitive to the strain induced in their arms by passing gravitational waves,
        and for certain detector geometries are sensitive to the polarization of the passing wave 
        \cite{Seto:2006dz,seto/taruya:2007, seto/taruya:2008, smith/caldwell:2016}. 
        
%--- what we did
  
  In this paper we seek to provide a realistic forecast of the ability of 
  LiteBIRD to distinguish this SU(2) model proposed in Ref.~\cite{dimastrogiovanni/etal:2016} from the conventional GWB generation by vacuum fluctuations. 
  LiteBIRD is a proposed CMB satellite mission with the primary science
  goal of detecting the GWB with $r < 10^{-3}$ \cite{Hazumi:2012aa,matsumura/etal:2014,2016JLTP..tmp..169M}.
  Therefore its sensitivity will be focused in 
  the lowest two hundred multipoles where the B-mode signal is both strong
  and relatively uncontaminated by gravitational lensing.  We exclude Stage 
  4 from the analysis as we found that the chirality signal is contained in the 
  multipole range $2 \lesssim \ell  \lesssim 30$. Since Stage 4 experiments 
  will have B-mode surveys over the range $\ell \gtrsim 30$ \cite{cmbs4}, 
  they will be ill-suited to constrain chirality, and we do not consider 
  them further. Ref. \cite{ferte/grain:2014} consider a simple model for 
  detecting primordial chirality using the CMB, and conclude that ground-based
  small-scale experiments are not well-suited for pursuing this signal. 
   We also considered a COrE-type experiment, the results of which we do not 
  include in our analysis, as they are similar to LiteBIRD due to the dominant 
  impact of large scale foreground residuals for both instruments. 
  In our analysis we include four contributions to the 
  uncertainty in a measurement of the chiral GWB: instrumental noise, 
  foreground residuals from the imperfect cleaning of multi-channel data, 
  gravitational lensing, and the joint self-calibration of the instrument's 
  polarimeter. This provides a robust assessment of LiteBIRD's capability 
  to detect primordial chirality. 
  
  On the other hand, laser interferometer gravitational wave observatories 
  are sensitive to the GWB today, and  provide 
  probes of much smaller scales: $k_{\rm CMB} \sim 10^
  {-4} \ \perMpc $ vs. $k_{\rm interf}\sim 10^{13} \ \perMpc$ \cite{garcia-bellido/etal:2016}.
  
  In the case of single-field slow-roll inflation the tensor spectrum 
  is expected to have a small red-tilt ($n_T=-r/8$, where 
  $n_T$ is the tilt of the tensor spectrum $P_h \sim k^{n_T}$), in which 
  case modern interferometers would not be sensitive enough to make a 
  detection. However, given 
  the scale-dependence of the model of Ref.~\cite{dimastrogiovanni/etal:2016} for part of the parameter space the 
  small scale tensor spectrum is comparatively large. For symmetry reasons
  the nominal designs of space-based gravitational interferometers 
  are insensitive to the circular polarization of gravitational waves. Since
  we are interested in constraining chirality we therefore consider 
  `advanced' stages of the nominal design of 
  LISA \cite{amaro-seoane/etal:2013,Bartolo:2016ami}, following the proposed designs of Ref.~\cite{smith/caldwell:2016} which provide equal sensitivity to both
  intensity and polarization of the GWB.
  In this paper we show that interferometers and CMB observations provide complementary probes at different scales of the axion-SU(2) 's primordial tensor spectrum. We then consider the sensitivities of two designs of an advanced stage LISA mission, and compare to constraints achieved using the CMB. \\
  
  In  \S \ref{sec:theory} we review the model proposed by Ref.~\cite{dimastrogiovanni/etal:2016} and its prediction for the GWB.
  In \S \ref{sec:cmb} we forecast the ability of a LiteBIRD-like 
  CMB satellite mission to detect the TB and EB correlations expected
  due to the chiral tensor spectrum, in the presence of foreground 
  contamination, gravitational lensing, instrument noise, and 
  simultaneous self-calibration of the telescope's polarimeter.
  In \S\ref{sec:interferometers} we analyse the sensitivity of 
  space-based gravitational interferometers to the chiral gravitational 
  background expected by this model.
  Finally, in \S \ref{sec:discussion} we summarize our findings and 
  discuss our conclusions.
  
 \section{Theory}
 \label{sec:theory}        
 
 In this section we will briefly review the axion-SU(2) model proposed in Ref. \cite{dimastrogiovanni/etal:2016}. The model is described by the following Lagrangian:
 \begin{equation}
 \mathcal{L}=\mathcal{L}_{\rm inflaton}+
 \frac{1}{2}(\partial_\mu \chi)^2 -\mu^4 \left[1+\cos \left(\frac{\chi}{f}\right)\right]
 -\frac{1}{4}F^a_{\mu\nu}F^{a\mu\nu} +\frac{\lambda}{4f}\chi F_{\mu\nu}\tilde{F}^{a\mu\nu},
 \end{equation}
 where $\mathcal{L}_{\rm inflaton}$ denotes the unspecified  inflaton sector which realizes inflation and the generation of the curvature perturbation compatible with the CMB observation, $\chi$ is a pseudo-scalar field (axion) with a cosine type potential, $\mu$ and $f$ are dimensionful parameters and $\lambda$
 is a dimensionless coupling constant between the axion and the gauge field. $F^a_{\mu\nu}\equiv \partial_\mu A_\nu^a-\partial_\nu A_\mu^a- g \epsilon^{abc} A^b_\mu A^c_\nu$ is the field strength of $SU(2)$ gauge field and $\tilde{F}^{a\mu\nu}\equiv \epsilon^{\mu\nu\rho\sigma}F^a_{\rho\sigma}/(2\sqrt{-g})$ is its dual. Here, $g$ is the self-coupling constant of the gauge field and $\epsilon^{abc}$ and $\epsilon^{\mu\nu\rho\sigma}$ are the completely asymmetric tensors, $\epsilon^{123}=\epsilon^{0123}=1$.

\begin{figure*}
	\centering
	\includegraphics[]{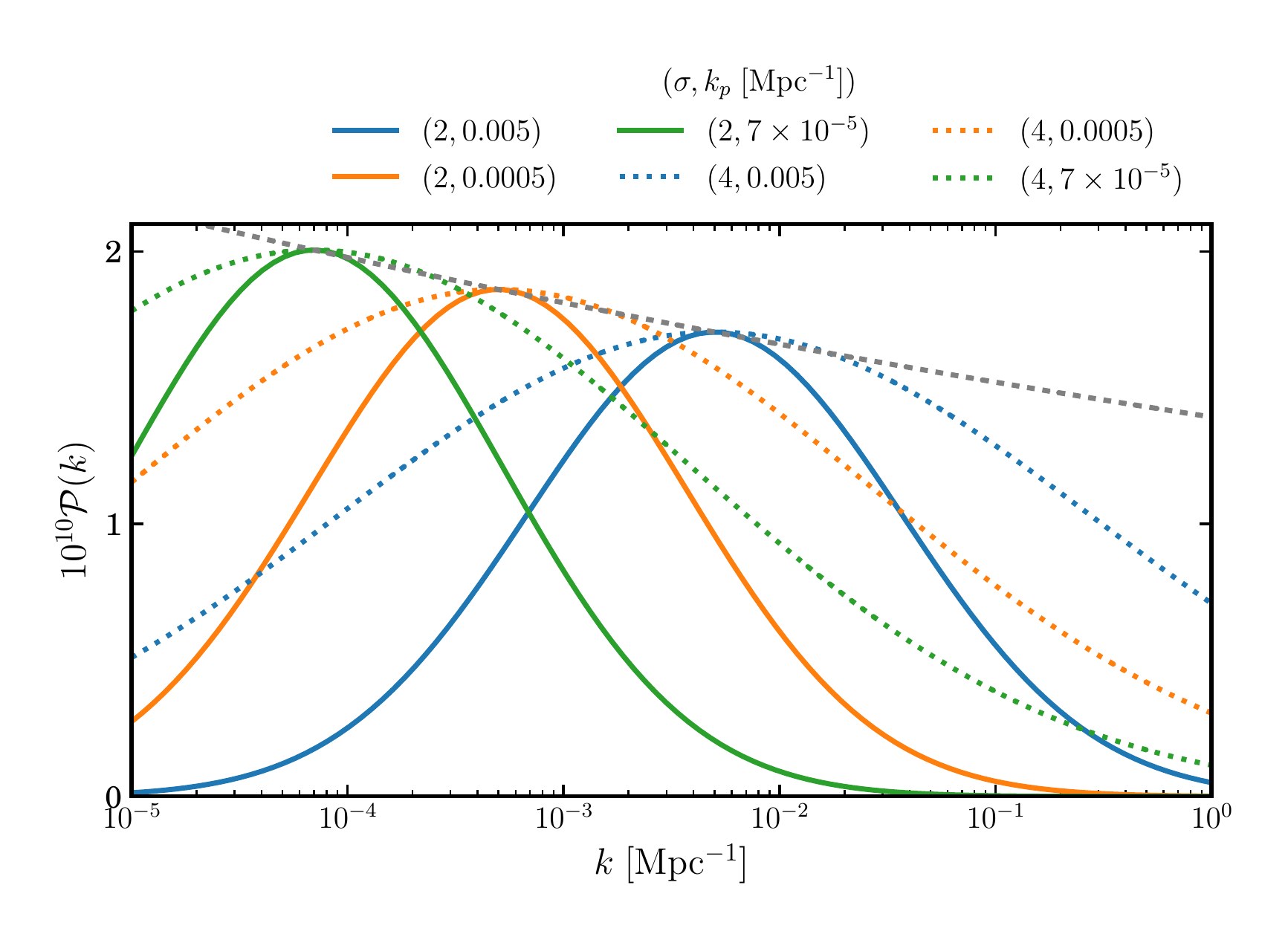}
	\caption{The predicted tensor spectrum, $\mathcal{P}^{\rm L,Sourced}_{h}$, for three sets 
		of parameters: (blue: $\sigma = 2, \ r_* = 0.07, \ k_p 
		= 0.005 \ {\rm Mpc^{-1}} \ $), (orange: $\sigma = 2, \ r_* 
		= 0.07, \ k_p = 0.0005 \ {\rm Mpc^{-1}} \ $), (green: $\sigma 
		= 2, \ r_* = 0.07, \ k_p = 7\times10^{-5} \ {\rm Mpc^{-1}}$).  }
	\label{fig:p_gw}
\end{figure*}

 In the axion-SU(2) model in the FLRW universe $g_{\mu\nu}={\rm diag}(1,-a^{2}(t),-a^{2}(t),-a^{2}(t))$, the SU(2) gauge fields naturally take an isotropic background configuration, $A^a_0=0,\ A^a_i=a(t)Q(t) \delta^a_i$ by virtue of the coupling to the axion $\chi$, and the transverse and traceless part of its perturbation, $t_{ij}=\delta A^i_j$ (not to be confused
 with the time variable, $t$), sources gravitational waves at the linear order. Interestingly, either of the two circular polarization modes of $t_{ij}$, namely $t_R$ or $t_L$, undergo a transient instability around the horizon crossing and gets substantially amplified. Subsequently, only the corresponding polarization mode of the gravitational wave, $h_R$ or $h_L$, is significantly sourced by $t_{ij}$ and fully chiral gravitational waves are generated. Note that the parity ($R\leftrightarrow L$) symmetry is spontaneously
 broken by the background evolution of the axion (i.e. the sign of $\partial_t \chi(t)$). 
In this paper we assume the left hand modes are produced for definiteness. In Appendix.~\ref{app:derivation}, we derive the following expression for the sourced GW power spectrum:
 \begin{equation}
 \label{eq:fitting_formula}
 \begin{aligned}
 \Pcal^{\rm L, \ Sourced}_h(k) &= r_{*}\Pcal_\zeta \exp \left[- 
 \frac{1}{2\sigma^2}\ln^2\left(\frac{k}{k_p}\right)\right] \\
 \Pcal^{\rm R, \ Sourced}_h(k) &\simeq 0,
 \end{aligned}
 \end{equation}
 where the amplitude is parameterized by the tensor-to-scalar ratio $r_*$ at the peak scale  $k=k_p$, $\sigma$ is the width of the Gaussian-shaped spectrum, and $\Pcal_\zeta$ is the power spectrum of curvature perturbations. We treat $r_*$ and $\sigma$ as free parameters in our analysis, while they can be rewritten in terms of more fundamental  parameters $m_*, \epsilon_{B*}$ and $\Delta N$, as discussed in Appendix~\ref{app:derivation}. Note that, there is no theoretical bound on $r_*$, while the possible values of $\sigma$ are restricted by $k_p$ as Eq.~\eqref{dN constraint}. Figure \ref{fig:CVebm} gives an example of how the amplitude 
 $r_*$ is degenerate in $m_*$ and $\epsilon_{B*}$, and we show an 
 example plot of $P_h^{\rm Sourced}(k)$ for three sets of these parameters in Figure \ref{fig:p_gw}.
 
 Here we define the power spectrum of primordial tensor perturbations to be:
 \begin{equation}
  \langle h^A_{\mathbf{k}} h^{A^\prime}_{\mathbf{k}^\prime} \rangle  =
 (2 \pi)^3 
 \frac{2 \pi^2}{k^3} \mathcal{P}_h^{A}(k) \delta^{(3)}(\mathbf{k} + 
 \mathbf{k}^\prime)\delta_{AA^\prime},
 \end{equation}
 where $A$ refers to the circular polarization of the gravitational wave with the momentum vector $\mathbf{k}:\ A = L, \ R$. 
 For the rest of this paper we model the primordial tensor spectrum as 
 being the sum of two contributions: a completely polarized sourced 
 contribution to the tensor spectrum %$\Pcal_h(k)$:
 $\Pcal_h^{\rm Sourced}$:
 and a contribution from the vacuum fluctuations, which we take to be 
 unpolarized and which we do not vary:
 
 \begin{equation}
 \begin{aligned}
 \mathcal{P}_h^{\rm vac} &= A_{\rm T} \left(\frac{k}{k_*} \right)^{
 n_{\rm T}} \\
 \mathcal{P}_{\zeta}^{\rm vac} &= A_{\rm S} \left(\frac{k}{k_*} \right)^{
 n_{\rm S}- 1}, \\
 \end{aligned}
 \end{equation}
  where $A_{\rm T} = r_{\rm vac}A_{\rm S}, \ A_{\rm S} = 2.2 \times 
 10^{-9},\ n_{\rm S} = 0.96, \ k_* = 0.05 \ {\rm Mpc}^{-1}$ are taken from the best-fit Planck cosmology \cite{planck_params:2016}. 
 We fix $r_{\rm vac} = 10^{-5}$ which corresponds to the inflationary Hubble scale $H_{\inf}=8\times 10^{11}$GeV and the tensor tilt is given by the consistency relation $n_{\rm T}=-r_{\rm vac}/8$. Note that $r_{\rm vac}$ is not required to be so small compared to 
 the sourced contribution; for larger values of $r_{\rm vac}$ the chiral contribution would be more difficult 
 to detect on the CMB due to the vacuum contribution to the BB spectrum. Therefore , we make the simplifying assumption of a small 
 $r_{\rm vac}$. In summary:
        \[
        \begin{aligned}
        \mathcal{P}_{\zeta}(k) &=\mathcal{P}_{\zeta}^{\rm vac} \\
        \mathcal{P}_h(k, k_p, r_*, \sigma) &=  \mathcal{P}^{\rm vac}_h(k) +  \mathcal{P}_h^{\rm 
                Sourced}(k, k_p, r_*, \sigma) \\
        \mathcal{P}^{\rm L}_h(k) - \mathcal{P}^{\rm R}_h(k)  &= \mathcal{P}^{\rm 
                Sourced}_h(k, k_p, r_*, \sigma).
        \end{aligned}
        \]
It is found  that contrary to the tensor perturbation, the scalar perturbations in the axion-SU(2) sector
do not have any instability for $m_Q\ge \sqrt{2}$ and they are even suppressed compared to
 the vacuum fluctuation of a massless scalar field due to their kinetic and mass mixing \cite{dimastrogiovanni/etal:2016, adshead/etal:2013, dimastrogiovanni/peloso:2013}. Since the axion-$SU(2)$ sector is decoupled from the inflaton and its energy density is subdominant, its contribution to the curvature perturbation is negligible. It is possible that the energy fraction of the axion $\Omega_\chi\equiv \rho_{\chi}/\rho_{\rm total}$ grows after inflation and $\chi$ becomes a curvaton if $\sigma$ is very large and the decay of the axion is suppressed more than that of the inflaton \cite{moroi/takahashi:2002, lyth/wands:2002, enqvist/sloth:2002}. In that case, the contribution from the scalar perturbations in the axion-SU(2) sector to the curvature perturbation may be significant and hence it would be interesting to investigate such cases. However, it is beyond the scope of this paper.
Therefore, we can simply consider that the curvature perturbation produced by the inflaton 
is not affected by the axion and the SU(2) gauge fields in this model. We may then take 
the TT, EE, and TE spectra to be given by constrained cosmological parameters (which we take to be: $h = 0.675 $, $\Omega_{\rm b} = 0.022$,
$\Omega_{\rm c} = 0.12$, $n_s = 0.96$, $\tau = 0.09$, $A_s = 2.2 \times 10^{-9}$), and investigate only the B-mode spectra: TB, EB, and BB.

\section{CMB}
\label{sec:cmb}

  In this section, we study the CMB phenomenology of the model 
  introduced in \S\ref{sec:theory}. The interesting CMB features of this 
   are the non-zero TB and EB spectra produced by the chiral
  tensor spectrum. We will calculate the expected TB and EB 
  spectra and make forecasts of their detectability by the LiteBIRD satellite
  in the presence of cosmic-variance, residual foregrounds, instrumental noise, 
  gravitational lensing, and polarimeter self-calibration.

\begin{figure*}
  \centering
  \includegraphics[]{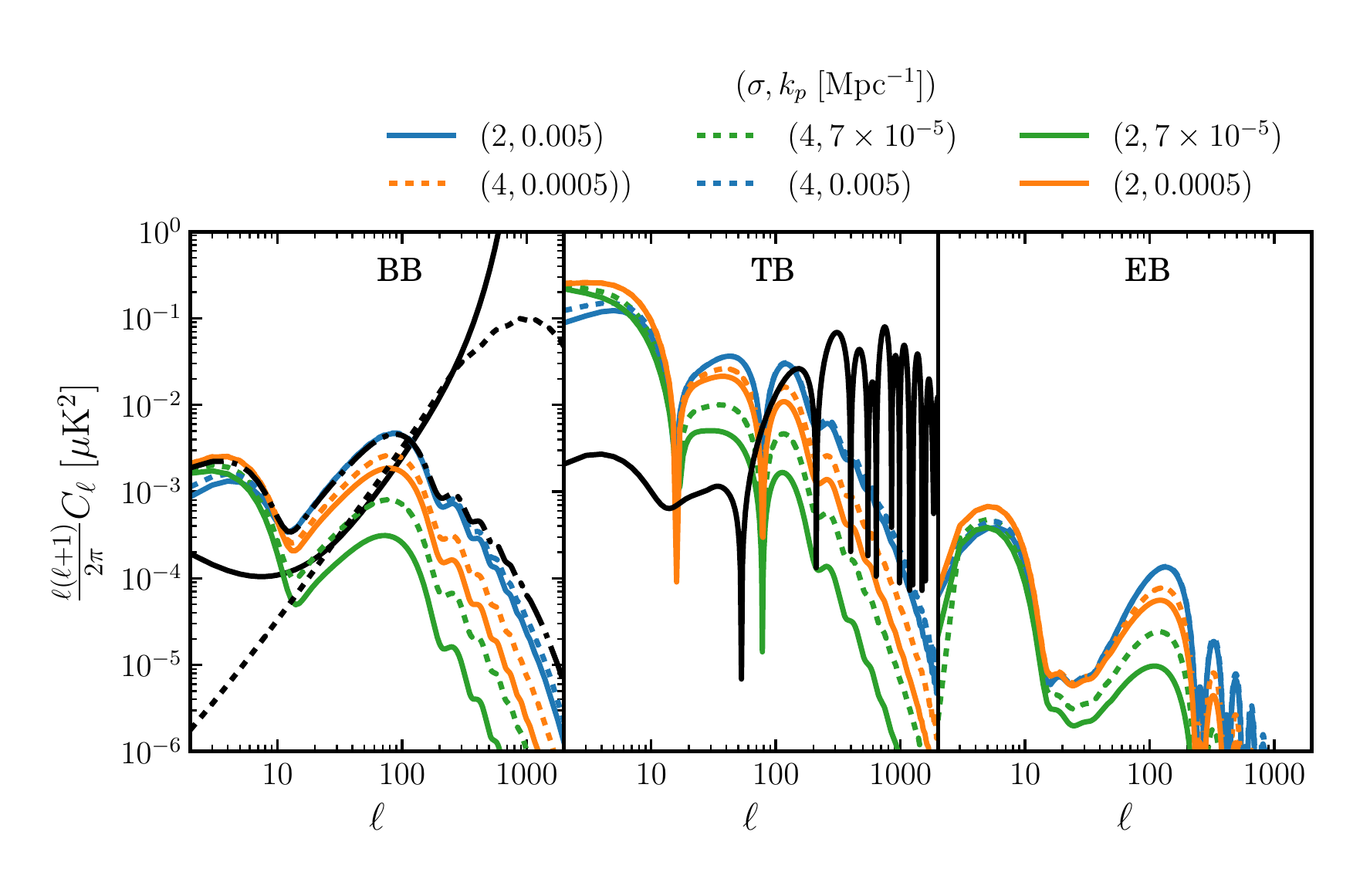}
  \caption{{\sc Left panel:} $C_\ell^{BB}$ for the same three 
  	sets of parameters used in Figure \ref{fig:p_gw}: (blue: $\sigma = 2, \ r_* = 0.07, \ k_p 
  = 0.005 \ \perMpc $), (orange: $\sigma = 2, \ r_* 
  = 0.07, \ k_p = 0.0005 \ \perMpc $), (green: $\sigma 
  = 2, \ r_* = 0.07, \ k_p = 7\times10^{-5} \ {\rm Mpc^
        {-1}}$) compared to the LiteBIRD noise spectrum 
  with 2\% foregrounds (solid black), the lensing BB spectrum (dashed black), and the 
  standard vacuum fluctuation $C^{\rm BB}_\ell(r=0.07)$ consistent 
  with the BKP $r<0.07 \ (95\% \ {\rm C.L.})$ (dash-dot black). The axion-SU(2)
  spectra contain  
  a contribution from vacuum fluctuations with $r=10^{-5}$, as is used in 
  the text. \\
  {\sc Right panel:} $|C_\ell^{TB}|$ (solid colour) and $|C_\ell^{EB}|$ 
  (dashed colour)  spectra for the same 
  three sets of parameters. Shown in black is an example of the spurious 
  TB signal induced by polarimeter
  miscalibration for an angle of one arcminute, as discussed in 
  \S \ref{subsec:simdet}.}
  \label{fig:signal_summary}
  \end{figure*}

  The anisotropies on the CMB are calculated by the integration of 
  the primordial perturbation spectra over the transfer functions 
  describing the evolution of perturbations with time. The 
  tensor contribution to the angular power spectra of the 
  anisotropies are \cite{Lue:1998mq,saito/etal:2007,glusevic/kamionkowski:2010,gerbino/etal:2016}
  \begin{equation}
  \label{eq:cmb_spectra}
  \begin{aligned}
  C_\ell^{t, X_1X_2} &= 4 \pi \int d(\ln k) \left[ \Pcal_h^{L}(k)
  +\Pcal_h^{R}(k) \right]\Delta^t_{X_1,\ell}(k)\Delta^t_{X_2 , 
  \ell}(k), \\
  C_\ell^{t, Y_1Y_2} &= 4 \pi \int d(\ln k) \left[ \Pcal_h^{L}(k)
  -\Pcal_h^{R}(k) \right]\Delta^t_{Y_1,\ell}(k)\Delta^t_{Y_2, 
  \ell}(k).
  \end{aligned}
  \end{equation}
  where $X_1X_2 = \{TT,\ TE,\ EE,\ BB \}$ and $Y_1Y_2 = \{ 
  TB,\ EB\}$, and $\Delta^t_{X,\ell}(k)$ indicates the tensor transfer function \cite{Pritchard:2004qp}. To calculate these spectra we use the {\sc CLASS} code 
  \cite{class}, making the necessary modifications for it to 
  calculate TB and EB spectra. In Figure
  \ref{fig:signal_summary} we plot examples of the BB and TB spectra calculated 
  in this way for a few different combinations of the model parameters, and 
  compare them to the noise contributions from lensing, instrument noise and 
  foreground residuals that we will consider later.
  
  In this paper we assess the detectability of the chirality of the primordial 
  GWB over the parameter space spanned by $(r_*, k_p, \sigma)$. Therefore, we 
  calculate the expected signal-to-noise of the combined TB and EB spectra 
  \cite{shiraishi/etal:2016}:
  \begin{equation}
  \label{eq:tbebsn}
   \left( \frac{S}{N} \right)^2_{\rm TB + EB} = 
   \sum_{\ell=2}^{\ell_{\rm max}} \sum_{X_1X_2, X_3X_4}  C_\ell^{X_1X_2}[\xi^{-1}]_
   \ell^{X_1X_2X_3X_4} C_\ell^{X_3X_4},
  \end{equation}
  where $X_1X_2, X_3X_4 = \{TT, EE, BB, TE, TB, EB\}$, and $\xi$ is the covariance 
  of our estimate of the power spectra given a certain theoretical and experimental 
  setup: $\xi^{X_1X_2X_3X_4} = \langle (\hat{C}^{X_1X_2}_\ell - 
  C^{X_1X_2}_\ell)(\hat{C}^{X_3X_4}_\ell - C^{X_3X_4}_\ell)  
  \rangle = \frac{1}{(2\ell+1)f_{\rm sky}}(\tilde C_\ell^{X_1X_3}\tilde C_
  \ell^{X_2X_4}+\tilde C_\ell^{X_1X_4}\tilde C_\ell^{X_2X_3})$,
  where tildes indicate the observed spectrum: $\tilde{C}^{XX^
  \prime}_\ell = C_\ell^{XX^\prime} + N_\ell^{XX^\prime}$, with $N_\ell^{XX^\prime}$ 
  denoting the noise spectrum, and the calculation of $\xi$ is detailed in Appendix 
  \ref{app:covariance_matrix}. $\ell_{\rm max}$ denotes the highest multipole 
  we consider, which in this case is 500. 
 
  Similarly, we can calculate the detectability of the primordial GWB, as opposed 
  to its chirality, by calculating the signal-to-noise of its contribution to the 
  BB spectrum. In the case of no lensing, this is simply:
  \begin{equation}
  \label{eq:bbsn}
  \left(\frac{S}{N}\right)_{\rm BB}^2
  = f_{\rm sky}  \sum_{\ell = 2}^{\ell_{\rm max}} 
  \frac{(2 \ell + 1)}{2}\left[ \frac{C^{BB}_\ell}
  {\tilde{C}^{BB}_\ell} \right]^2,
  \end{equation}
  
  However, one of the major sources of uncertainty in a measurement of the 
  BB spectrum is due to gravitational lensing. As the CMB propagates to 
  us from the surface of last scattering it is gravitationally lensed 
  by the intervening matter density, converting primary E-mode anisotropies to 
  secondary B-mode anisotropies, which then need to be accounted for in 
  measurements of BB \cite{zaldarriaga/seljak:1998}. 
  
  We can separate the contributions to BB into $C^{BB}_\ell = C_\ell^{BB, 
  \ {\rm Prim}}+C^{BB, \ {\rm Lens}}$, where `Prim', `Lens' refer to the 
  primordial and lensed contributions respectively. We are interested in measuring  
  $C_\ell^{BB, \ {\rm Prim}}$, and in effect $C^{BB, \ {\rm Lens}}$ 
  acts as an extra source of noise, with an unknown amplitude. 	The modification required to 
  calculate the signal-to-noise of the primordial BB signal is to consider the 
  $2\times2$ matrix: 
      \begin{equation}
      \mathcal{F}_{ij} = \sum_{\ell = 2}^{\ell_{\rm 
      max}} \frac{(2\ell + 1)f_{\rm sky}}{2} \frac{C^{BB, \ i}_
      \ell C^{BB, \ j}_\ell}{(\tilde{C}_\ell^{BB})^2},
      \end{equation}
      such that:
      \begin{equation}
       \label{eq:BBprimsn}
      \left(\frac{S}{N}\right)^2_{BB , i} = \frac{1}{(\mathcal{F}^{-1}
      )_{ii}}
      \end{equation}
      where the indices $i, \ j$ run pver `Prim', `Lens'. 
  Note that we will assume that the temperature spectrum is
  perfectly known over the range of scales we are interested 
  in, and that the sourced contribution to the scalar spectrum 
  is negligible \cite{dimastrogiovanni/etal:2016}: $\tilde{C}_\ell
  ^{\rm TT} = C_\ell^{\rm TT}$. 

  \subsection{Cosmic-variance limited case}
  \label{seubsec:cv}
\begin{figure*}
    \centering
    \includegraphics[]{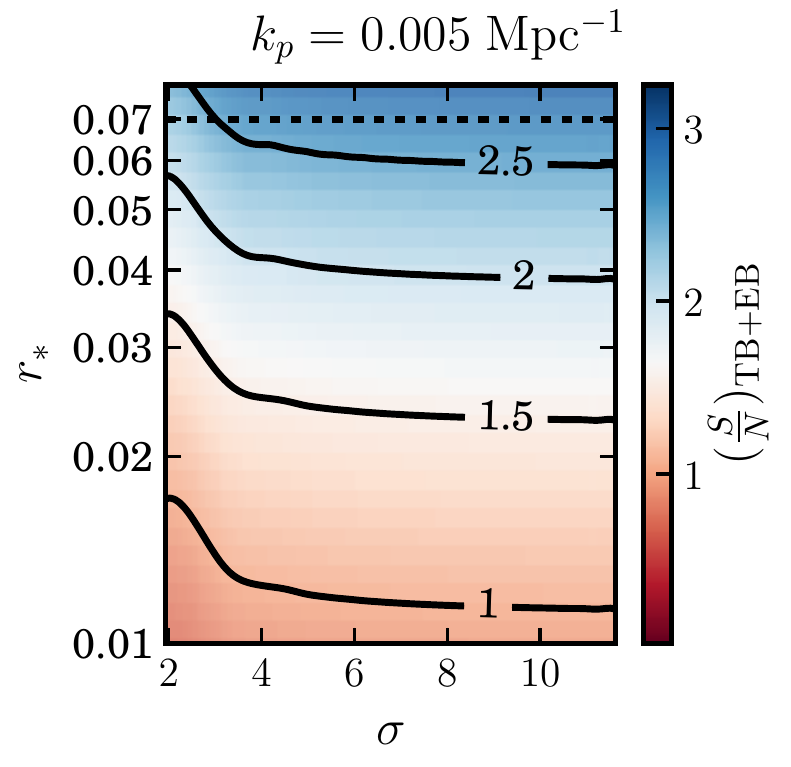}
    \includegraphics[]{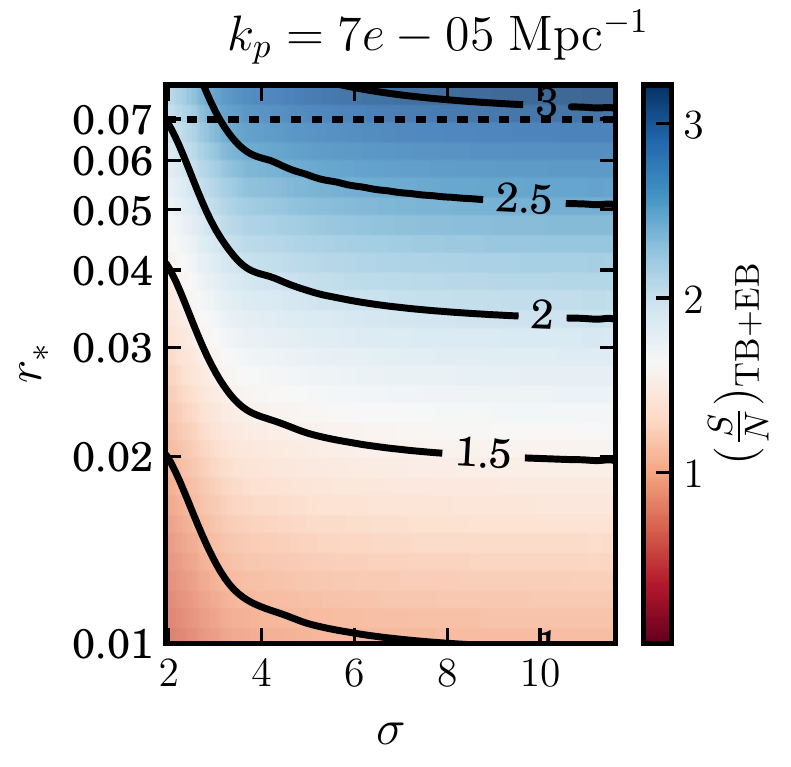}
    \caption{Signal-to-noise of TB + EB spectra assuming the perfect case 
    	of $f_{\rm sky} = 1$, with no foreground contamination, noiseless observations, and no delensing. The black dashed line indicates the 
    	bounds placed by 
    $r_* < 0.07$. {\sc Left panel:} $k_p = 5 \times 10^{-3} \perMpc$. 
    {\sc Right panel:}  $k_p = 7 \times 10^{-5} \perMpc$.}
    \label{fig:TBEB_cv}
    \end{figure*}

    Here, we discuss the signal-to-noise of the TB, EB, and BB 
    spectra in the case of cosmic variance-limited observations:
    $\tilde{C}^{XX^\prime}_\ell = C_\ell^{XX^\prime}$.
    In this scenario, in the absence of lensing, Equation 
    \ref{eq:bbsn} has the simple analytic form 
    $\left(\frac{S}{N}\right)_{\rm BB}^2 =  f_{\rm sky}(
    \ell_{\rm max} + 3)(\ell_{\rm max} - 1) / 2$.
    The signal-to-noise of the TB and EB spectra calculated using 
    Equation \ref{eq:tbebsn} are shown in 
    Figure \ref{fig:TBEB_cv} for the parameter space of the 
    model, assuming a lensed BB spectrum with $f_{\rm sky}=1$. We 
    consider only $r_* < 0.07$, in line with 
    current observational constraints on the scale-invariant tensor-to-scalar
    ratio $r_{0.05}<0.07 \ (95 \% {\rm C. L.})$, where the subscript indicates the pivot scale
    in $\perMpc$ \cite{bkp16}. Figure \ref{fig:signal_summary} demonstrates
    that the shape of $C_\ell^{\rm BB}$ is strongly dependent on the 
    position of the peak in the GW spectrum, $k_p$, and also on the 
    width of the peak, $\sigma$. Therefore,  
    the BKP bound on $r$ does not simply imply the same bound on $r_*$; 
    a small value of $k_p$ and $\sigma$ could allow a 
    large value of $r_*$ without exceeding the BKP limit, due 
    to the small scale damping of $C^{\rm BB}_\ell$. However, excepting 
    $\mathcal{O}(1)$ underestimation for small $k_p$, the 
    BKP bound provides a useful guide as to what is allowed by 
    current observations.
    
    The values of $k_p = 7 \times 10^{-5} \perMpc$
    and $k_p= 5 \times 10^{-3} \perMpc$ were chosen as they probe 
    different scales to which the CMB is sensitive.
    $\Pcal_h^{\rm L, \ Sourced}(k)$ is more sharply peaked for 
    smaller $\sigma$ and so for a given $r_*$ 
    the signal-to-noise decreases with decreasing $\sigma$.
    As $\sigma$ increases the tensor spectrum becomes
    almost scale-invariant over the range of scales accessible with the 
    CMB and so the signal-to-noise does not depend on $\sigma$ for large values
    of $\sigma$. Figure \ref{fig:TBEB_cv} 
    shows that the maximum achievable signal-to-noise is $\sim 3$
    and that the chirality is undetectable with $\frac{S}{N} 
        \lesssim 1$ for $r_*\lesssim0.01$.
        
  \subsection{Including instrument noise and foreground contamination}
  \label{subsec:fgns_lensed}

    We now consider instrument noise, contamination of the spectrum due to imperfect
    foreground separation, and assume that we are unable to perform 
    any `delensing'. \\
    
    The model we use for the noise spectrum includes the instrument 
    noise in the CMB channels, the residual foregrounds in the final 
    CMB map (assumed to be at a level 
    of 2 \%, following Refs. \cite{matsumura/etal:2014, katayama/komatsu:2011, shiraishi/etal:2016, oyama/etal:2016}) and the 
    instrumental noise from channels used for 
    foreground cleaning that is introduced into the CMB channels by 
    the cleaning process. The details of how we combine 
    these factors to produce a final noise contribution to the measured
    CMB spectrum, as well as the instrument specifications for LiteBIRD
    can be found in Appendix \ref{app:aggregate_noise}. 
    In the left panel of Figure \ref{fig:signal_summary} we show the 
    contributions to the BB noise spectrum, $N_\ell^{\rm BB}$, from lensing, LiteBIRD instrumental 
    noise, and foreground residuals compared to the primordial 
    $C_\ell^{\rm BB}$. 
    
    \subsubsection{BB Signal-to-Noise}
    \label{subsec:lensedbbsn}

\begin{figure*}
      \centering
      \begin{subfigure}
      	\centering
      	\includegraphics[]{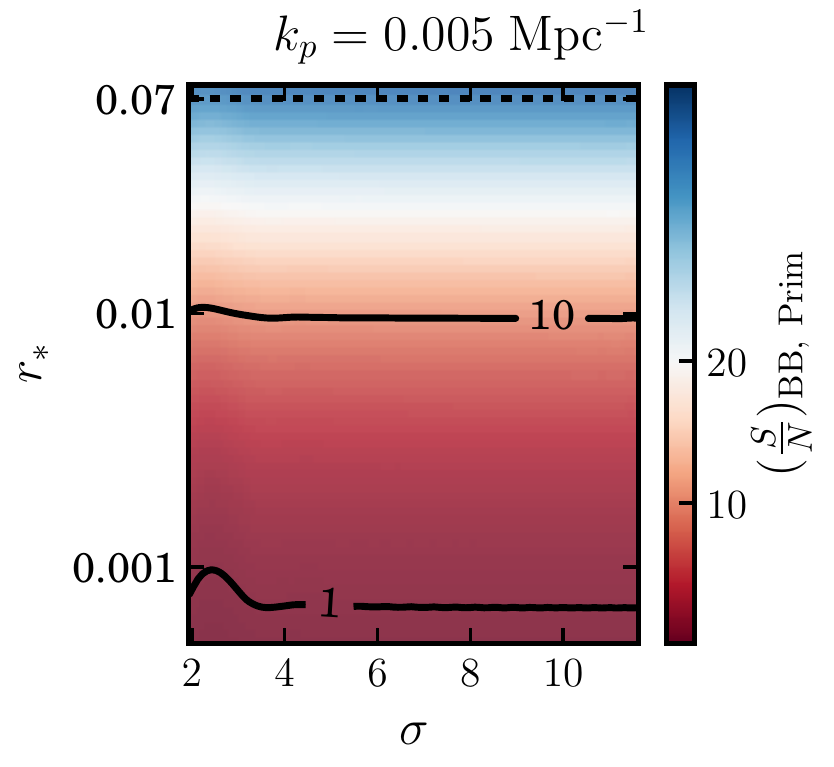}
      \end{subfigure}%
	  \begin{subfigure}
	  	\centering
	  	\includegraphics[]{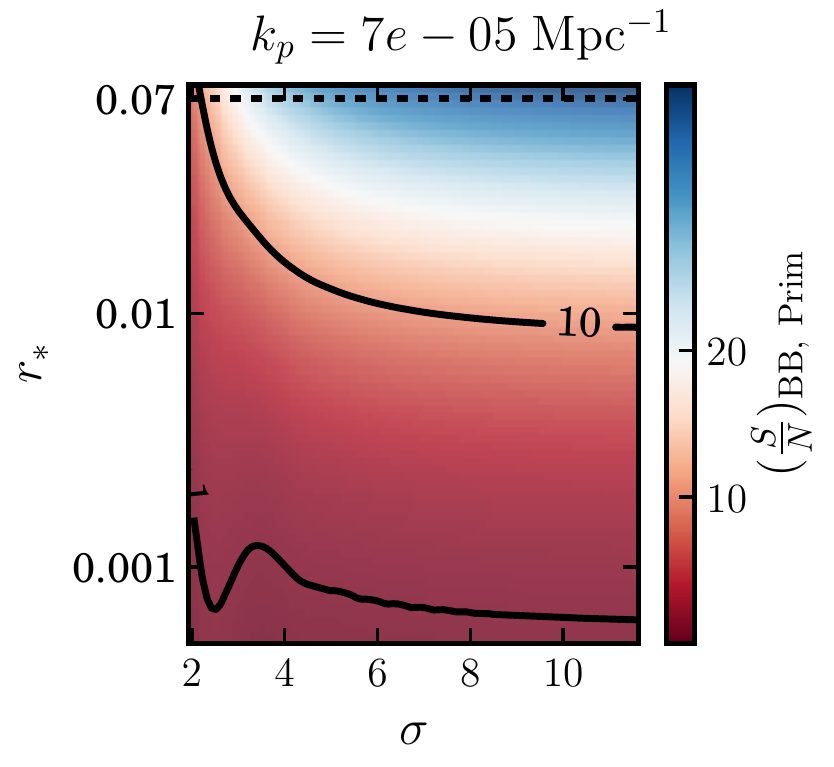}
	  \end{subfigure}%
      \caption{Signal-to-noise of BB spectrum assuming 
      no delensing and 2\% foreground contamination and LiteBIRD 
      instrumental noise added using method described in Appendix 
      \ref{app:aggregate_noise}. The dashed line refers
      to the observational constraint of $r_*=0.07$. The signal-to-noise 
      achieved in BB is much larger than that in TB+EB as the cosmic 
      variances of BB and TB are proportional to $(C^{BB}_\ell)^2$, and 
      $C_\ell^{BB}C_\ell^{TT}$, respectively. The factor of 
      $C_\ell^{TT}$ means cosmic variance in the TB spectrum is much more 
      significant than in BB.  {\sc Left panel:} $k_p = 5 \times 10^{-3} \perMpc$. {\sc Right panel:} $k_p = 7 \times 10^{-5} \perMpc$}
      \label{fig:BBprim_fglens_sn}
      \end{figure*}

   	 We calculate Equation \ref{eq:BBprimsn} over the available parameter 
   	 space and show the result in Figure \ref{fig:BBprim_fglens_sn}. In a 
     similar way to the TB and EB signal-to-noise we see that there is some dependence
     on $\sigma$, especially in the case of smaller $k_p$. This is expected since
     $k_p =  7 \times 10^{-5} \perMpc$ is slightly smaller than those scales 
     to which we expect the CMB to be sensitive \cite{garcia-bellido/etal:2016}. Therefore,
     we expect that reducing $\sigma$ for this value of $k_p$ will eventually exclude 
     the tensor perturbations from contributing to CMB scales, explaining the 
     sharp decrease in S/N for low $\sigma$ and a given $r_*$. From 
     Figure \ref{fig:BBprim_fglens_sn} it is clear that we can detect
     the primordial contribution to BB for $r_* > 10^{-3}$, which is 
     consistent with the aim of LiteBIRD to achieve an uncertainty on 
     the null case of $r=0$ of less than $10^{-3}$.

    \subsubsection{TB+EB Signal-to-Noise}
    \label{subsec:lensedtbsn}

\begin{figure*}
      \centering
      \includegraphics[]{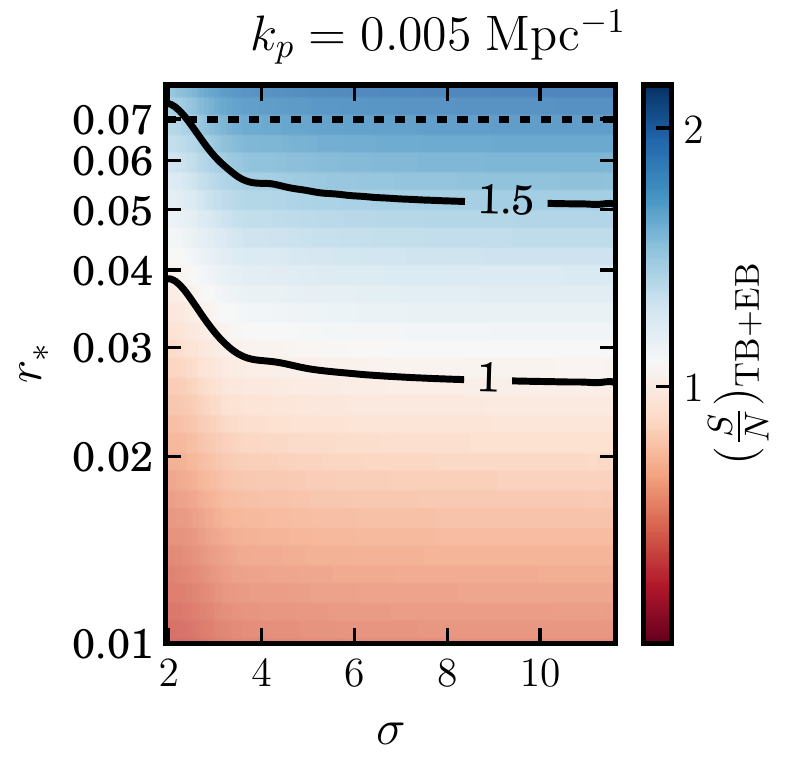}
      \includegraphics[]{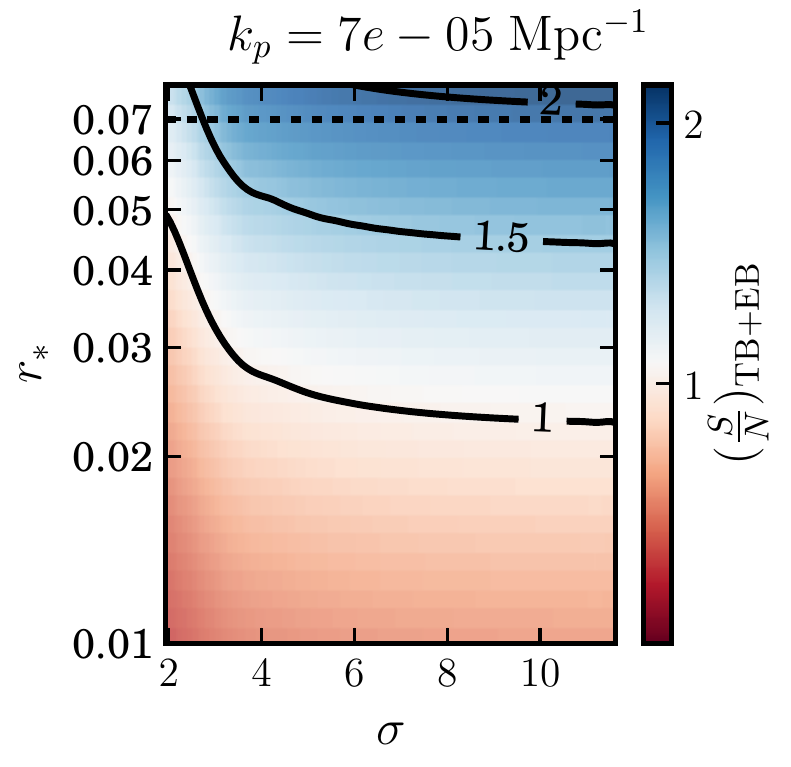}
      \caption{Signal-to-noise of TB + EB spectra 
      assuming no delensing and 2\% foreground contamination and 
      LiteBIRD instrumental noise added using method described 
      in Appendix \ref{app:aggregate_noise}. The dashed line refers
      to the observational constraint of $r_*=0.07$. {\sc Left panel} 
      $k_p = 5 \times 10^{-3} \perMpc$. {\sc Right panel:} $k_p = 7 \times 
      10^{-5} \perMpc$.}
      \label{fig:TBEB_fglens_sn}
      \end{figure*}

      Lensing affects the TB and EB signal-to-noise only through  
      $\tilde{C}_\ell^{BB}$, since the direct lensing contributions 
      to TB and EB are negligible \cite{saito/etal:2007,shiraishi/etal:2016}. We calculate
      Equation \ref{eq:tbebsn} over the available parameter space, 
      now including instrument noise for a LiteBIRD-type experiment
      (with parameters shown in Table \ref{tab:lbird_specs}), foreground residuals, and gravitational 
      lensing, and show the result in Figure \ref{fig:TBEB_fglens_sn}. 
      Over the allowed parameter space, the maximum achievable 
      signal-to-noise is $\frac{S}{N}\sim 2$. Whilst for $r_* 
      \lesssim 0.03$ LiteBIRD can not detect  chirality in this model, compared to $r_* 
      \lesssim 0.01$ in the CV-limited case. The right panel of Figure
          \ref{fig:signal_summary} demonstrates that the TB and EB signal peaks at 
          $\ell \lesssim 10$, making the large scale foreground residual contribution 
          to the noise, shown in the left panel of Figure \ref{fig:signal_summary}, the 
          dominant factor causing this reduction in sensitivity. 
      
      Improvements in foreground cleaning algorithms could reduce the level 
      of foreground contamination, and perhaps allow a larger sky fraction to be 
      used in the analysis. However, even with perfect control of these factors, the 
      cosmic-variance limit of Figure \ref{fig:TBEB_cv} can not be beaten.  We conclude 
      from this study that the most important factor limiting the sensitivity of CMB 
      observations to the chirality of the GWB is the large cosmic variance of the 
      TB and EB spectra due to large scalar T and E signals, respectively. 
      
  \subsection{Simultaneous detection and self-calibration}
	\label{subsec:simdet}
        In order to achieve its baseline performance target, 
        LiteBIRD will require an uncertainty on the polarimeter calibration angle of 
        less than one arcminute \cite{odea/etal:2007, shimon/etal:2008}.
        There are several methods that have been used in the past 
        to calibrate polarimeters such as polarized astrophysical 
        sources like the Crab Nebula (Tau A), or
        man-made sources such as a polarization selective mesh.
        There are many factors preventing such methods achieving calibrations
        better than one degree.  
        For example, Tau A is the best candidate for a point-like polarized
        source, but this provides a calibration uncertainty of $\sim0.5$ degrees 
        \cite{kaufman/etal:2014}, and with these it is hard to achieve a calibration uncertainty 
        better than one degree \cite{planck_calibration:2016}. The polarization 
        of Tau A also has a poorly understood frequency dependence, and is 
        ultimately an extended source, making it poorly suited to a characterization
        of the polarized beam \cite{nati/etal:2017}.
        Man-made sources on the other hand must often be placed in the near 
        field and are unstable over long time frames. However, a recent 
        proposal of a balloon-borne artificial polarization source 
        in the far field of ground-based experiments may ameliorate
        this problem for ground-based telescopes \cite{nati/etal:2017}.  \\
    
        LiteBIRD plans to self-calibrate its polarimeter using the EB spectrum, which 
        is assumed to have zero contribution from primordial perturbations 
        \cite{keating/etal:2013}. Unfortunately this makes assumptions about cosmology, and uses part
        of the constraining power to calibrate the instrument, instead of for 
        science. Furthermore, residual foreground 
        contributions to TB and EB may result in a biasing of the calibration
        angle. Ref. \cite{abitbol/etal:2016} shows that a miscalibration 
        angle of 0.5 degrees can result in a bias in the recovered
        value of $r$ of $2 \times 10^{-3}$, which is significant for 
        LiteBIRD's aim to push constraints on $r$ to $r\sim 10^{-3}$. 
        However, Ref. \cite{abitbol/etal:2016} also finds that TB and EB 
        are consistent with zero in a study of the low-foreground 
        BICEP2 region. Furthermore, in a study of the Planck data 
        Ref. \cite{planck_dust:2016} finds that 
        TB and EB are both consistent with zero for sky fractions up to 
        $f_{\rm sky} = 0.3$, and that TB increases to significant 
        levels only for larger sky fractions, whilst EB is only marginally
        non-zero for $f_{\rm sky} = 0.7$. Therefore whilst foregrounds must be 
        considered, they do not necessarily limit the use of this approach to 
        calibration. \\
        
        We want to study the detectability of primordial TB and EB correlations when taking self-calibration into account. The self-calibration process is carried 
        out by zeroing the miscalibration $\Delta \psi$ by measuring its contribution 
        to the TB and EB spectra. In this analysis we will assume that residual foreground contributions to TB and EB are negligible. 
        
        If the angle of the polarimeter is miscalibrated by some angle 
        $\Delta \psi$ the measured $Q, \ U$ will be rotated. We work 
        with the spin-2 quantities $(Q \pm iU)(\hat n)$ which have the 
        transformation properties under rotation:
    \[
    (\tilde Q \pm i\tilde U)(\hat n) = 
      %% e^{i \pm 2\Delta \psi}
   e^{\pm i 2\Delta \psi}  (Q 
    \pm iU)(\hat n).
    \]
    E and B can be computed to find:
    \[
    \begin{pmatrix}
      \tilde a^T_{\ell m} \\
      \tilde a^E_{\ell m} \\
      \tilde a^B_{\ell m} \\
    \end{pmatrix}
    = 
    \begin{pmatrix}
      1 & 0 & 0 \\
      0 & \cos(2\dpsi) & -\sin(2\dpsi) \\
      0 & \sin(2\dpsi) & \cos(2 \dpsi)
    \end{pmatrix}
    \begin{pmatrix}
      a^T_{\ell m} \\
      a^E_{\ell m} \\
      a^B_{\ell m} \\
    \end{pmatrix}
    \]
    which give the resulting rotations of the angular power spectra:
    \begin{equation}
    \label{eq:rotated_cls}
    \begin{pmatrix}
       C_\ell^{TE} \\
       C_\ell^{TB} \\
       C_\ell^{EE} \\
       C_\ell^{BB} \\
       C_\ell^{EB}  
    \end{pmatrix}_{\rm rot}
    = 
    \begin{pmatrix}
      \cos(2\dpsi) & -\sin(2\dpsi) & 0 & 0 & 0 \\
      \sin(2\dpsi) & \cos(2 \dpsi) & 0 & 0 & 0 \\
      0 & 0 & \cos^2(2\dpsi) & \sin^2(2\dpsi) & -\sin(4\dpsi) \\
      0 & 0 & \sin^2(2\dpsi) & \cos^2(2\dpsi) & \sin(4\dpsi) \\
      0 & 0 & \frac{\sin(4\dpsi)}{2} & -\frac{\sin(4 \dpsi)}{2} & 
      \cos(4\dpsi)
    \end{pmatrix}
    \begin{pmatrix}
      C_\ell^{TE} \\
      C_\ell^{TB} \\
      C_\ell^{EE} \\
      C_\ell^{BB} \\
      C_\ell^{EB}
    \end{pmatrix}.
    \end{equation}
    We then replace the primordial spectra in our expression 
    for $\tilde C_\ell$ with the rotated spectra:
    \[
    \tilde C^{XX^\prime}_\ell =  C_{{\rm rot} \, \ell}^{XX^\prime} + N^
    {XX^\prime}_\ell.
    \]
    We jointly estimate the uncertainty on the miscalibration angle
    and the recovered amplitude of the TB and EB spectra parametrized
    by $r_*$ using the Fisher information:
    \begin{equation}
     \label{eq:polcalsn}
    \mathcal{F}_{ij} = \sum_{X_1X_2,X_3X_4} \sum_{\ell = 2}^{\ell_{\rm max}} 
    \frac{\partial C^{X_1X_2}_\ell}{\partial a_i}
    [\xi^{-1}]^{X_1X_2X_3X_4} \frac{\partial C^{X_3X_4}_\ell}{\partial
    a_j}
    \end{equation}
    where %$i,j = \dpsi, r_*$
    $a_i, a_j = \dpsi, r_*$
    . The uncertainty on the miscalibration 
    angle is then given by $(\sigma_{\dpsi})^2 = (\mathcal{F}^{-1}
    )_{\dpsi \dpsi}$. We can easily calculate the derivatives 
    with respect to $\dpsi$ in Equation \ref{eq:polcalsn} using 
    Equation \ref{eq:rotated_cls}. In order to calculate the 
    derivatives with respect to $r_*$ we write:
    $C^{TB/EB}_\ell = r_*C^{TB/EB}_\ell(r_* = 1)$. In order to study 
    the interaction of the miscalibration angle and primordial 
    chirality we calculate the correlation coefficient:
    \[
    \alpha \equiv \frac{\mathcal{F}_{\dpsi r_*}}{\sqrt{\mathcal{F}
                _{\dpsi \dpsi}\mathcal{F}_{r_*r_*}}}.
    \]
    
\begin{figure*}
    	\begin{subfigure}
    		\centering
    		\includegraphics[width=.48\linewidth]{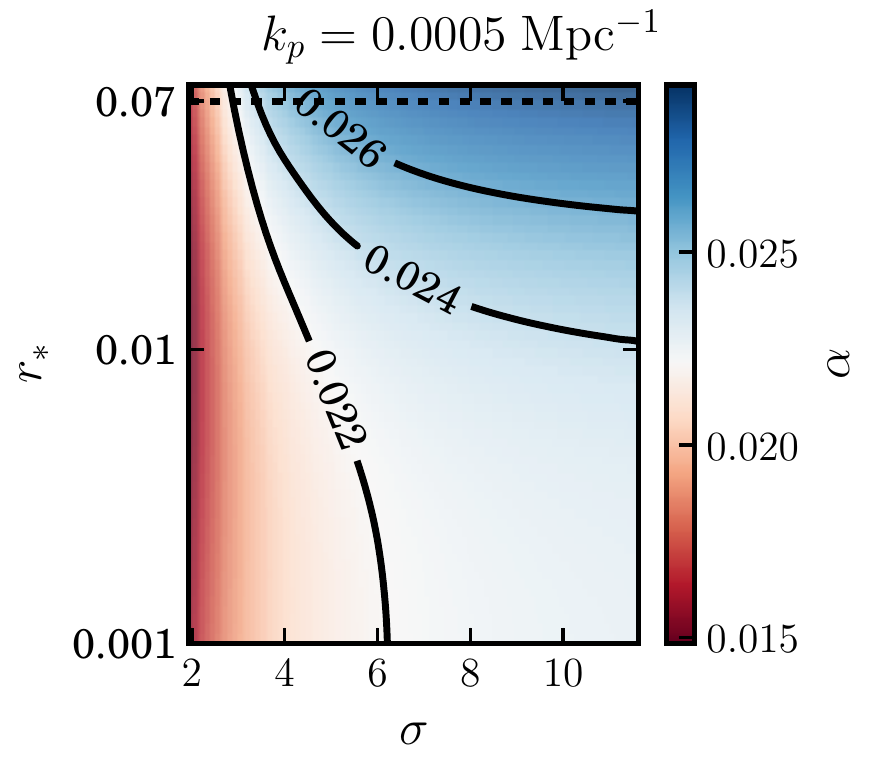}
    	\end{subfigure}%
    	\begin{subfigure}
    		\centering
    		\includegraphics[width=.48\linewidth]{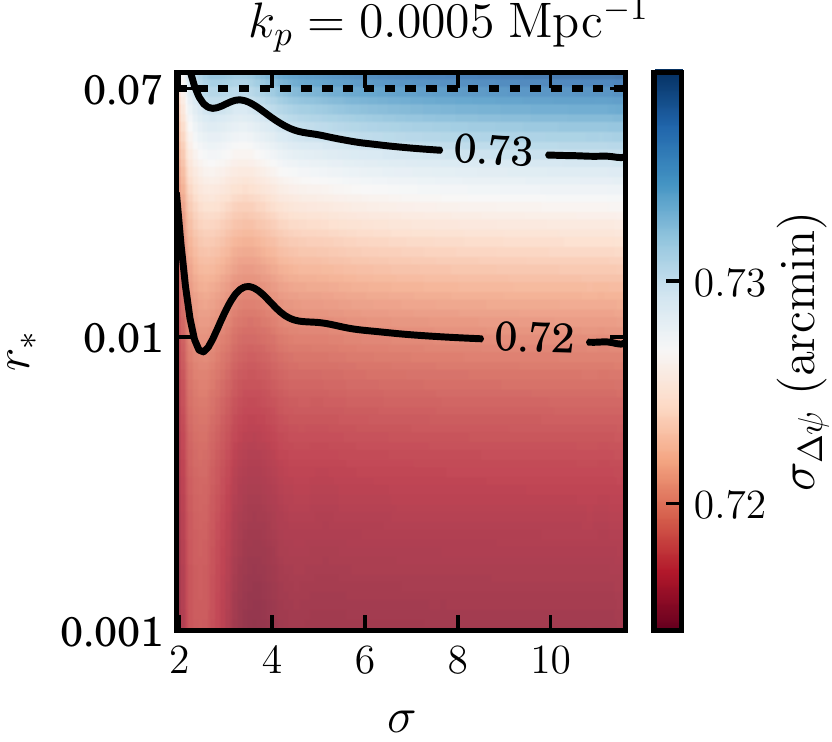}
    	\end{subfigure}
    \caption{Correlation coefficient $\alpha$ (left) and 1-$\sigma$ uncertainty on the polarimeter calibration (right) for LiteBIRD,
    calculated over the allowed parameter space of the model assuming 2\% foreground residual and no delensing. The dashed line shows 
    the observational constraint of $r_* = 0.07$.}  
    \label{fig:corrco}
    \end{figure*}

    We now calculate the 1-sigma uncertainty in 
    a measurement of the miscalibration angle $\dpsi$ and $\alpha$ 
    over the allowed parameter space of the model and show the resulting 
    contour plots in Figure \ref{fig:corrco}.  \\ 
    
    We find that for LiteBIRD $\sigma_{\dpsi} < 1$ arcmin for 
    all of the allowed space, making the simultaneous 
    calibration of the polarimeter and detection of the 
    parity-violation possible. The correlation coefficient is less 
    than 0.03 for the allowed parameter space, indicating that the effects of primordial parity-violation and miscalibration are easily separable. 
    This can be understood from the right panel of 
    Figure \ref{fig:signal_summary} where it is clear that the 
    primordial signal is a large scale effect, with maximum signal
    at $\ell \sim 10$, whereas the contribution to TB from 
    miscalibration is a small scale effect which dominates at 
    $\ell > 100$. This is supported by the $\sigma$ dependence of
    $\alpha$ in the left panel of Figure \ref{fig:corrco}. The 
    two effects become more correlated for larger values of 
    $\sigma$ which correspond to flatter spectra, and hence more power at small scales.
    Varying $k_p$ has little effect on the result that the 
    effects are separable, but does introduce some interesting 
    dependence on $\sigma$. This indicates that a sufficiently high 
    $\ell_{\rm max}$ is necessary for the separation of these
    effects. For smaller values of $k_p$, $\sigma_{\Delta \psi}$ 
    becomes more dependent on $\sigma$. For example, with $k_p = 7 
    \times 10^{-5} \perMpc$, for a given $r_*$, $\sigma_{\Delta \psi}$ 
    increases with $\sigma$ since the flatter spectra of large $\sigma$ become more important when $k_p$ is further away from the small scales at which 
    the miscalibration effect occurs. On the other hand when $k_p = 5 \times 10^{-3} \perMpc$ 
    the dependence on $\sigma$ is reversed. This is because the miscalibartion
    effect peaks at $\ell \sim 100$, which corresponds to contributions from
    modes around $k\sim 100 / \eta_0 = 7 \times 10^{-3} \perMpc$ ,
    where $\eta_0$ is the comoving distance to the surface of last scattering. 
    Therefore an increase of $\sigma$ for $k_p \sim 7\times 10^{-3} \perMpc$
    will make the signals less correlated as the flatter spectra will 
    introduce more power at larger scales.
    
    In conclusion, any reduction in sensitivity to TB and EB due to the 
    calibration requirements is negligible, and is ignored in the results 
    we quote for LiteBIRD. Our results are in agreement with Refs \cite{ferte/grain:2014, glusevic/kamionkowski:2010}, which also find that the primordial and 
    miscalibration contributions are readily separable.
  
  \subsection{CMB Results}
  \label{sec:cmbresults}
    Here, we summarize the findings of the CMB section and provide a 
    prognosis of the usefulness of CMB observations in detecting 
    gravitational wave chirality.
    
    In the case of cosmic variance-limited ultimate observations we found
    that over the parameter space of the model the maximum signal-to-noise achievable was $\sim3$ for the largest values of $r_*$,
    and that the chirality is undetectable for $r_* \lesssim 0.01$, 
    in agreement with previous studies of simpler models of chiral 
    GWBs with nearly scale-invariant spectra 
    \cite{glusevic/kamionkowski:2010,gerbino/etal:2016,saito/etal:2007}. 
    Moving on to the realistic case of a LiteBIRD-like experiment with 
    no delensing capability, a 2\% level of foreground residuals, 
    and a simultaneous self-calibration,
    we find that for the largest allowable values of $r_*$ it 
    may achieve a signal-to-noise of 2.0, making the chirality
    detectable. The chirality is undetectable by LiteBIRD 
    for $r_* \lesssim 0.03$. \\

    Though a detection with a two sigma significance may be of interest, it
    is only achievable for a small part of the parameter space, $0.03 \lesssim 
    r_* \lesssim 0.07$, and in any event we have demonstrated that we may not
    exceed a $\frac{S}{N}$ of 3 using CMB two-point statistics. We also investigated
    a COrE+ design with the same level of foreground residuals as LiteBIRD and found 
    that is performed very similarly to LiteBIRD since both instruments would be 
    limited by foreground residuals on the large scales we are interested 
    in. As stated in \S\ref{sec:intro} we will not gain anything extra from Stage 4 
    observations, as they are limited to $\ell\gtrsim 30$. 
    Therefore, in order to make stronger statistical detections of this model
    using the CMB, higher order statistical techniques taking advantage of the 
    model's non-Gaussianity may have more success as shown for the axion-U(1) model \cite{shiraishi/etal:2016}. \\
    
    Alternatively, we can investigate different physical probes altogether. 
    In the next section, we consider complementary constraints on the axion-SU(2) 
    model from space-based laser interferometer gravitational wave observatories. 
    
    \begin{figure*}
    \centering
    \includegraphics[]{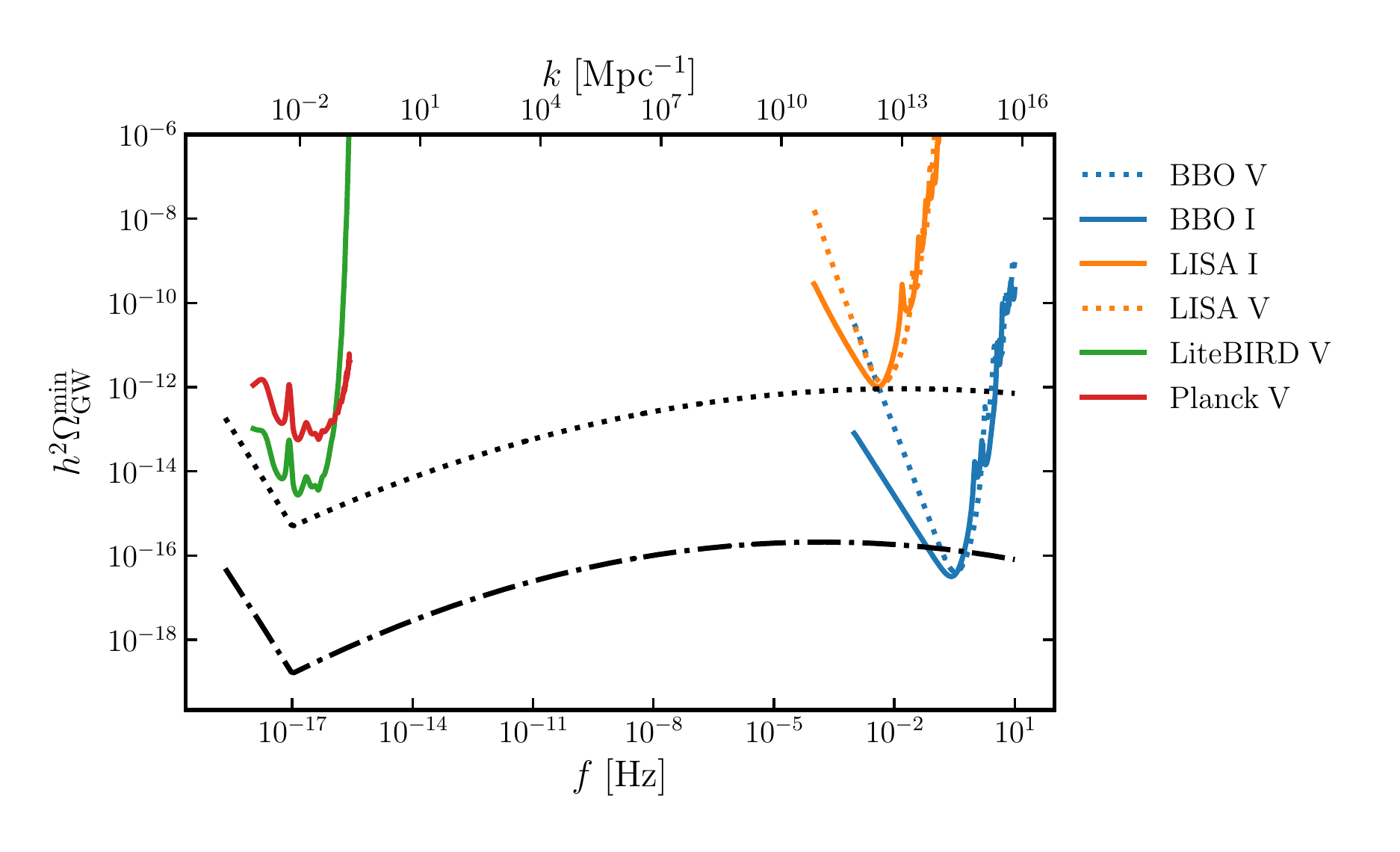}
    \caption{Comparison of the sensitivity curves for LiteBIRD, 
    Planck, LISA, and BBO corresponding to a signal-to-noise of one 
    at a given frequency in intensity ($I$) or polarization ($V$). At the top 
        horizontal axis we also show the corresponding wavenumber computed via 
        $\frac{k}{\rm Mpc^{-1}} = 6.5 \times 10^{14} \frac{f}{\rm Hz} $. 
    Also plotted are the primordial spectra for the 
        parameters: $k_p = 10^{13} \perMpc$, $\sigma = 9$, $r_* = 835$ (black dotted) and 
        $k_p = 10^{11} \perMpc$, $\sigma = 8 $, $r_* = 0.15$ (black dash-dotted). Note that below $f=10^{-17} \ {\rm Hz}$ transfer
        function of the fractional gravitational wave background energy density 
        changes due to the transition between matter and radiation 
        dominated eras.
    We see that even for the large values of $\sigma$ required by the large $k_p$ values of 
    the axion-SU(2) model LISA and BBO can make a detection that would 
    still be inaccessible at CMB scales. This motivates the 
    evaluation of signal-to-noise for the interferometers. 
    Note that the CMB sensitivity curves have been smoothed with 
    a Gaussian kernel due to the sharp oscillations introduced 
    by the transfer function (Equation 
    \ref{eq:cmb_transfers}).}
    \label{fig:comparison}
    \end{figure*}

\section{Laser Interferometers}
\label{sec:interferometers}

        Due to the strong scale-dependence of the tensor spectrum,
        it may be possible to study the case of large $k_p$ 
        using laser interferometer gravitational wave observatories. 
        Previous studies have indicated that the scale-invariant
        spectrum of single-field slow-roll inflation would be 
        too weak at interferometer scales to be detected
        by current generation interferometers such as LIGO \cite{LIGO:2009}, 
        VIRGO \cite{VIRGO:2012}, and LISA \cite{Bartolo:2016ami}. However, the model we 
        consider has a large feature at $k_p$, therefore for $k_p 
        \sim 10^{11}-10^{13} \ \text{Mpc}^{-1}$, current generation 
        interferometers may be sensitive to the GWB of the axion-SU(2) model. 
        
        It should be noted that it is difficult to have a sourced gravitational 
        wave spectrum with a sharp peak on interferometer scales. 
        This is because of the attractor behaviour of the background 
        axion field coupled to the SU(2) gauge fields (see Appendix
        \ref{app:derivation}). As a result, we consider the rather flat 
        spectra  seen in Figure \ref{fig:comparison}. For such flat
        spectra one may expect any signal detectable with interferometers
        would also be detectable on CMB scales, making the use of 
        interferometers redundant. We therefore first demonstrate 
        the complementarity of our CMB and interferometer studies.
        We compare their sensitivities as a function of the frequency
        $f$ of the gravitational wave background. 
        The quantity we use to compare sensitivities is the minimum 
        detectable fractional  energy density in primordial gravitational 
        waves today: 
        \begin{equation}
        \label{eq:Odef}
        \Omega_{\rm GW}(f) \equiv \frac{1}{\rho_c}\frac{\partial 
                \rho_{\rm GW}}{\partial \ln(f)}
        \end{equation}
        where $\rho_c$ is the critical density to close the Universe 
        evaluated today, and $\rho_{\rm GW} = \frac{c^2}{32\pi^2 G}
        \langle \dot h_{ij} \dot h_{ij} \rangle$, where $h_{ij} \equiv \delta g_{ij}^{TT} / a^2$. The calculation
        for the CMB is detailed in Appendix \ref{app:cmb_freqdep}, and 
        for interferometers in the remainder of this section. 
        Figure \ref{fig:comparison} displays the minimum detectable 
        fractional energy density using the CMB and interferometers 
        for Planck, LiteBIRD, an advanced LISA \cite{Bartolo:2016ami}and 
        BBO \cite{crowder/cornish:2005}. We see that LiteBIRD has a much 
        improved sensitivity to chirality, compared to Planck, which is
        due to its much lower instrumental noise.
        The two plotted theoretical spectra are clearly detectable by LISA or BBO, 
        without being detectable at CMB scales, making interferometers
        an independent, complementary probe of the primordial spectrum of 
        the axion-SU(2) model. 

 \subsection{Interferometer notation}

  Laser interferometers consist of a set of test masses 
  placed at nodes and linked by laser beams. 
  Interferometry is used to measure
  the change in the optical path length between 
  test masses. A passing gravitational wave induces a time-dependent oscillation in the optical path length, which 
  can be isolated from noise by taking cross-correlations 
  between detectors. 
 
  The metric perturbation at point $\bx$ at time $t$, $h_{ij}(t, \bx)$, can be decomposed into a 
  superposition of plane waves \cite{allen:1997}:
  \[
    h_{ij}(t, \bx) = \sum_P \int d^3 \bk C_P(\bk) \sin(
    ckt-\bk \cdot \bx + \Phi(k))e_{ij}^P(\vO),
  \]
  where we use the transverse traceless basis tensors with 
  normalization $e_{ij}^P(\vO)e_{ij}^{P^\prime}(\vO) = 
  2\delta_{PP^\prime}$, and $P = +, \times$. It is more 
  convenient to deal with complex values, and so we rewrite 
  this as:
  \[
  h_{ij}(t, \bx) = \sum_P \fint \Oint \ h_P(f, \vO) 
  \exp(2\pi i f(t - \frac{\bx \cdot \vO}{c})) e^P_{ij}(\vO),
  \]
  where $ck = 2 \pi f$, $\bk \cdot \bx = 2 \pi f \frac{\vO 
  \cdot \bx}{c}$, and $\vO$ is a unit vector in the 
direction of propagation of the gravitational wave. Since the coefficients satisfy $h_P(f, \vO) 
= h^*_P(-f, \vO)$, $h_{ij}(t, \bx)$ is explicitly real. 
 The theory we are dealing with produces a highly 
        non-Gaussian GWB \cite{agrawal/etal:2017}.
 We can summarize the two-point statistics using the 
 following expectation values of the Fourier coefficients,
 but this will not capture all the available information: 
 \begin{equation}
 \label{eq:polarized2point}
 \begin{aligned}
 \langle h_P(f, \vO) h^*_{P\pr}(f\pr, \vO\pr) \rangle &= \frac{1}{2} \delta(f-f\pr)\frac{\delta^{(2)}(\vO - \vO\pr)}{4 \pi}S^{PP\pr}_h(f)\\
 \begin{pmatrix}
 \langle h_+(f, \vO)h^*_+(f^\prime, \vO^\prime) \rangle & 
 \langle  h_+(f, \vO)h^*_\times(f^\prime, \vO^\prime) 
 \rangle \\
 \langle h_\times(f, \vO)h^*_+(f^\prime, \vO^\prime) 
 \rangle & \langle  h_\times(f, \vO)h^*_\times(f^\prime, 
 \vO^\prime) \rangle
 \end{pmatrix} 
 &= \frac{1}{2}\delta(f - f^\prime)\frac{\delta^{(2)}(\vO 
        - \vO^\prime)}{4\pi}
 \begin{pmatrix}
 I(f) & iV(f) \\
 -iV(f) & I(f) 
 \end{pmatrix},
 \end{aligned}
 \end{equation}
 where $I(f)$ and $V(f)$ are the Stokes parameters for intensity and circular polarization respectively. As shown below, $V(f)$ quantifies the difference between the amplitudes of two circular polarization states and hence is a clean observable for the chiral GWB \cite{Seto:2006dz,seto/taruya:2007,seto/taruya:2008}.

\begin{figure*}
\centering
\includegraphics[]{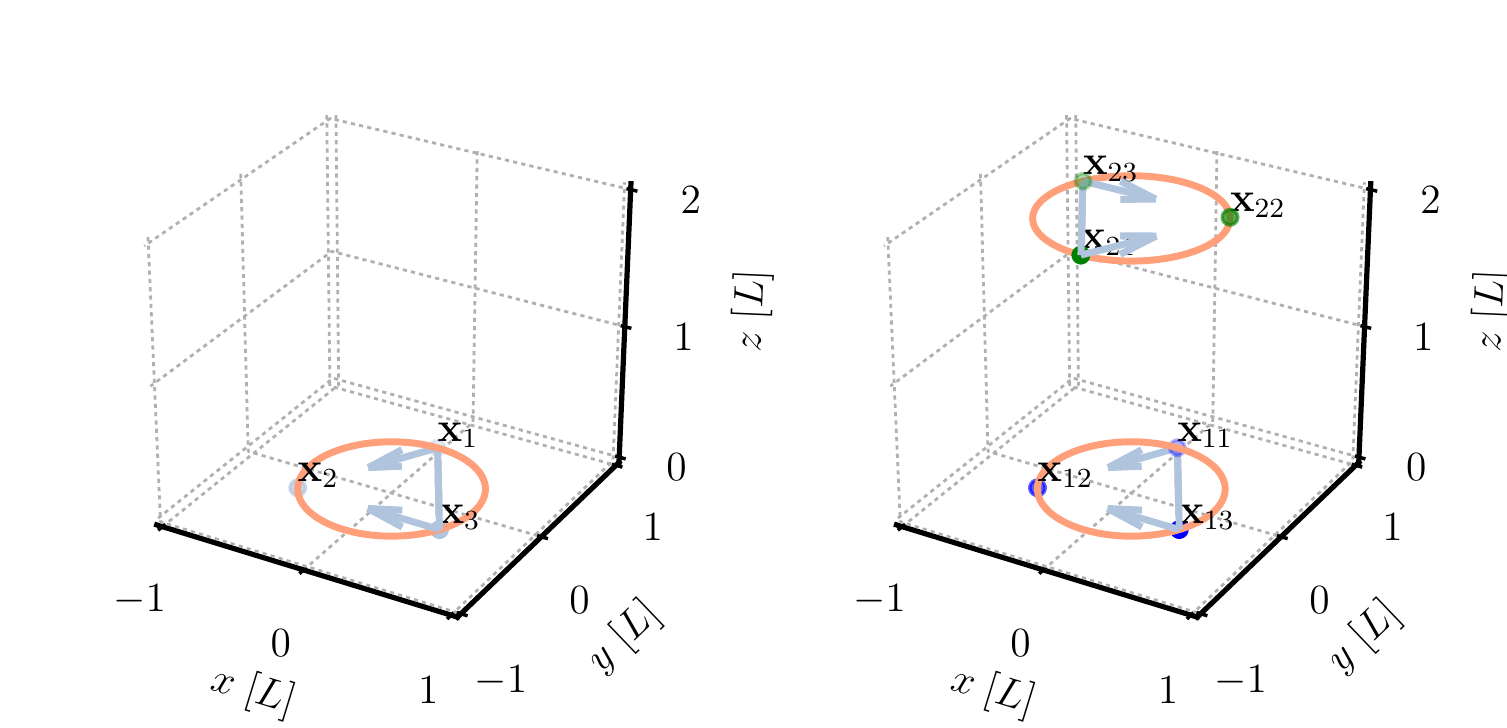}
\caption{Possible designs for future space-based laser interferometers.
 The blue arrows show the laser links used in the 
Michelson interferometer signals defined in Equation \ref{eq:michelson}.
{\sc Left panel:} One constellation design of a 
space-based interferometer, this corresponds to the baseline 
LISA design. The points $\bx_{i}$ show the $i^{\rm th}$ satellite.
{\sc Right panel:} An advanced stage design of LISA
or BBO with two constellations. The points $\bx_{ij}$ show the $j^{\rm th}$ satellite on the $i^{\rm th}$ constellation.}
\label{fig:interferometer_setup}
\end{figure*}

\subsection{Interferometer response}
\label{subsec:interferometer_response}

        In this section, we present the design of the interferometers
        for which we will forecast the sensitivity to a polarized 
        gravitational wave background. This analysis uses the 
        designs proposed by Ref. \cite{smith/caldwell:2016}. We summarise 
        some of the results of Ref. \cite{smith/caldwell:2016} here, however for 
        further details we refer readers to Ref. \cite{smith/caldwell:2016}. 

        Let us consider a set of masses placed at positions 
        $\bx_i$, and the phase change, $\Dphi_{ij}$,
        of light as it travels from mass $i$ at time 
        $t_i$ to mass $j$ arriving at time $t$ \cite{finn:2009}:
        \begin{equation}
        \label{eq:dphi}
        \Delta \phi_{ij}(t) =  \int_{-\infty}^\infty df \int d^2 
        \hat{n} \sum_P \tilde{h}_P (f, \hat{n}) e^P_{ab}(\hat n) 
        e^{i2 \pi f (t_i - \hat{n} \cdot \mathbf{x}_i)} D^{ab}(
        \hat{u}_{ij}\cdot \hat{n}, f),
        \end{equation}
        where $D^{ab}$ is the single-arm transfer function which
        contains all the geometric information about the instrument
        and must be derived individually for each interferometer 
        set-up \cite{cornish:2001}, and $\hat{u}_{ij}$ is a unit vector 
        pointing from detector $i$ to detector $j$. We now define
        the Fourier transform of a signal $g(t)$ observed for a 
        time  $T$: $g(f) = \int_{T/2}^{T/2}g(t) \exp(-2\pi i ft)$. The 
        Fourier transform of the phase change $\Dphi$ is then:
        \begin{equation}
        \begin{split}
        \Dphi_{ij}(f) &= \int_{-T/2}^{T/2} dt \fint^\prime \Oint \sum_P 
        h_P(f^\prime, \vO)\exp \left(i2\pi f^\prime(t-\frac{\bx 
        \cdot \vO}{c})-2\pi i ft)\right)D^{ab}(\hat{u}_{ij}\cdot 
        \hat{n}, f^\prime) \\
        &= \fint^\prime \delta_T(f-f^\prime) \Oint \sum_P 
        h_P(f^\prime, \vO)\exp \left(-i2\pi f^\prime\frac{\bx 
                \cdot \vO}{c}\right)D^{ab}(\hat{u}_{ij}\cdot 
        \hat{n}, f^\prime),
        \end{split}
        \end{equation}
        where $\delta_T$ is a finite-time approximation to 
        the delta function defined as $\delta_T(f-f^\prime) \equiv 
        T {\rm sinc}\left( \pi T(f-f^\prime)\right)$, with the 
        properties: $\delta_T(0) = T$, $\lim_{T\to\infty} \delta_T(f) 
        \rightarrow \delta(f)$. We may 
        form a signal by constructing a linear combination of phase 
        changes along different paths around the instrument, and 
        then cross-correlating these signals. The signal we seek to measure is stochastic and so to 
        distinguish it from noise we must cross-correlate the 
        detector output with the output from a detector with
        independent noise properties. The expectation of 
        the cross correlated signal will be composed of terms 
        like:
        \begin{equation}
                \begin{split}
                        \langle \Dphi_{ij}(f_1) \Dphi_{kl}(f_2) \rangle = 
                        &\fint\pr \fint\pp \Oint_1 \Oint_2 
                        \sum_{P_1 P_2} \delta_T(f_1 - f\pr)\delta_T(f_2 - 
                        f\pp) \langle h_{P_1}(f\pr, \vO_1) h_{P_2}(f\pp, \vO_2) 
                        \rangle \\
                        &\exp \left(-2 \pi if\pr t \vO_1 \cdot \bx_i \right)
                        \exp \left(-2 \pi if\pp t \vO_2 \cdot \bx_k \right)
                        D^{ab}(\hat{u}_{ij}\cdot \vO_1, f\pr)
                        D^{cd}(\hat{u}_{kl}\cdot \vO_2, f\pp)  
                        e^{P_1}_{ab}(\vO_1)e^{P_2}_{cd}(\vO_2) .
                \end{split}
        \end{equation}
        Using $\langle h_{P_1}(f\pr, \vO_1) h_{P_2}(f\pp, \vO_2)  
        \rangle =\langle h_{P_1}(f\pr, \vO_1) h^*_{P_2}(-f\pp, \vO_2)  
        \rangle$, and $D^{ab}(\hat{u}_{ij}\cdot \vO, -f) = 
        D^{ab*}(\hat{u}_{ij}\cdot \vO, f)$ we can write this as:
        \begin{equation}
                \begin{aligned}
                \langle \Dphi_{ij}(f_1) \Dphi_{kl}(f_2) \rangle &= 
                \frac{1}{2} \fint\pr \delta_T(f_1 - f\pr)\delta_T(f_2 - 
                f\pr)S_h^{P_1P_2}(f\pr)\Rcal^{ijkl}_{P_1P_2}(f\pr), \\
                \Rcal^{ijkl}_{P_1P_2}(f) &= \frac{1}{4 \pi}\Oint
                \exp \left(-2 \pi if \vO \cdot (\bx_i - \bx_k) \right)
                D^{ab}(\hat{u}_{ij}\cdot \vO, f)
                D^{cd}(\hat{u}_{kl}\cdot \vO, f)  
                e^{P_1}_{ab}(\vO_1)e^{P_2}_{cd}(\vO).
                \end{aligned}
        \end{equation}
        $\Rcal^{ijkl}_{P_1P_2}$ is referred to as the response function of 
        the detector. $\Rcal$ depends on the relative position and 
        orientation of the arms $i\rightarrow j$ and $k \rightarrow l$,
        as well as the transfer functions of the two arms. 
        
        In the remainder of this section we consider two interferometer
        designs. In \S\ref{subsubsec:one_constellation} we consider the baseline design for near-future space-based
        interferometers such as the European Space Agency-led Laser 
        Interferometer Space Antenna (LISA) \cite{amaro-seoane/etal:2013}, and in \S\ref{subsubsec:two_constellation} we consider two futuristic `advanced stage' LISA-like missions similar to the proposed Big Bang Observatory (BBO) \cite{crowder/cornish:2005}.
        
        \subsubsection{One constellation}
        \label{subsubsec:one_constellation}
        
        In this section, we consider the design shown in the left panel of Figure \ref{fig:interferometer_setup}. This is the baseline design of the LISA mission, and consists of three satellites placed at the vertices $\bx_i$ of an equilateral triangle of side $L$, and a total of six laser links between the satellites, allowing for measurement of the phase change $\Dphi_{ij}$ where $i, j = \{1, 2, 3\}, \ i \neq j$. We define the following three signals:
        \begin{equation}
        \label{eq:michelson}
        \begin{aligned}
        s^\alpha(t) &= \frac{1}{2} \left( \Delta \phi_{12}
        (t-2L) + \Delta \phi_{21}(t-L) - \Delta \phi_{13}(t-2L) - 
        \Delta \phi_{31}(t-L) \right) + n^\alpha(t), \\
        s^\gamma(t) &= \frac{1}{2}\left( \Delta \phi_{31}
        (t-2L) + \Delta \phi_{13}(t-L) - \Delta \phi_{32}(t-2L) - 
        \Delta \phi_{23}(t-L) \right) + n^\gamma(t), \\
        s^\beta(t) &= s^\alpha(t) + 2 s^\gamma(t).
        \end{aligned}
        \end{equation}
        The equilateral design means that the laser phase noise, 
        which is the dominant contribution
        to the noise terms $n(t)$, cancels \cite{cornish:2001}. 
        Furthermore Ref.~\cite{smith/caldwell:2016} shows that signals 
        $\alpha$ and $\beta$ have independent noise properties. 
        We therefore consider their cross-correlations:
        \begin{equation}
        \label{eq:basic_crosscorr}
        \langle s^{X_1}(f)s^{X_2}(f\pr) \rangle = \frac{1}{2}\delta(f-f^\prime) \left[\Rcal^{X_1X_2}_I(f)I(f) + \Rcal^{X_1X_2}_V(f)V(f) \right],
        \end{equation}
        where $X_1, X_2 = \{\alpha, \beta\}$, and:
        \begin{equation}
        \label{eq:RIV}
        \begin{aligned}
        \Rcal^{X_1X_2}_I(f) &= \frac{1}{4\pi} \Oint \left[ F_{X_1}^+
        (f,\hat{u} \cdot \vO)F_{X_2}^{+*}(f, \hat{u} \cdot \vO) + 
        F_{X_1}^\times(f,\hat{u} \cdot \vO)F_{X_2}^{\times*}(f, 
        \hat{u} \cdot \vO) \right], \\
        \Rcal^{X_1X_2}_V(f) &= \frac{1}{4\pi} \Oint \left[ F_{X_1}^+
        (f,\hat{u} \cdot \vO)F_{X_2}^{\times*}(f, \hat{u} \cdot 
        \vO) - F_{X_1}^\times(f,\hat{u} \cdot \vO)F_{X_2}^{+*}(f, 
        \hat{u} \cdot \vO) \right] ,
        \end{aligned}
        \end{equation}
        and
        \cite{cornish:2001, smith/caldwell:2016, romano/cornish:2016}:
        \begin{equation}
        \begin{aligned}
        F^P_X(f, \vO \cdot \hat{u}) &= D^X_{ij}(f, \hat{u} \cdot 
        \vO)e^P_{ij}(\vO), \\
        D^{\alpha}(f, \hat{u} \cdot \vO) 
        &= \frac{1}{2}\exp(
        -2\pi i f \vO \cdot \mathbf{x}_{1}) \left[ \hat{u} 
        \otimes \hat{u} \ T_{\rm MI}(f, \hat{n} \cdot \hat{u}) 
        - \hat{v} \otimes \hat{v} \ T_{\rm MI}(f, \vO \cdot 
        \hat{v}) \right], \\
        D^{\beta}(f, \hat{u} \cdot \vO) 
        &= D^\alpha(f, \hat{u} \cdot \vO) + \exp(-2\pi i f 
        \hat{n}\cdot \mathbf{x}_{3}) \left[ \hat{u} \otimes 
        \hat{u} \ T_{\rm MI}(f, \vO \cdot -\hat{u}) - 
        \hat{w} \otimes \hat{w} \ T_{\rm MI}(f, \vO 
        \cdot - \hat{w}) \right], \\
        T_{\rm MI}(f, \vO \cdot \hat{u}) &= \frac{1}{2}\left[
        {\rm sinc}\left(\frac{f(1 - \hat{u} \cdot \vO)}{2 
        f_*}\right){\rm exp}\left(-i \frac{f}{2f_*}(3 + \hat{u}
        \cdot \vO)\right) + {\rm sinc}\left(\frac{f(1+\hat{u}
        \cdot\vO)}{2f_*}\right){\rm exp}\left(-i \frac{f}{2f_
        *}(1 + \hat{u} \cdot \vO)\right)\right]. \nonumber
        \end{aligned}
        \end{equation}
        
        Consider the instrument's response to a gravitational wave 
        travelling in the direction $\vO = (\theta, \phi)$, and 
        another travelling in a direction 
        with $\theta \rightarrow \pi - \theta$, i.e. reflected 
        in the $x-y$ plane. Since the vectors $\hat{u}, \hat{v},
        \hat{w}, \bx_i$ are all in the $x-y$ plane it is easy to 
        see that the products $\bx \cdot \vO, \hat{u} \cdot \vO$
        etc. are invariant. Under this transformation only the
        $z$ part of the basis tensor $e^+_{ab}(\vO)$ is altered. 
        Since $D^{ab}(f, \vO \cdot \hat{u})$ is non-zero only 
        in the $x-y$ part, then the product $D^{ab}(f, \vO \cdot \hat{u})e_{ab}^+(\vO)$ is invariant. On the other
        hand the $x-y$  part of the $e_{ab}^\times(\vO)$ tensor changes 
        sign, meaning that $D^{ab}(f, \vO \cdot \hat{u})e_{ab}
        ^\times(\vO)$ changes sign. Therefore, when performing 
        the angular integral
        in Equation \ref{eq:RIV} the terms with a single power of 
        $F^\times_X(f, \vO \cdot \hat{u})$ go to zero, giving 
        $\Rcal_V^{X_1X_2}(f) = 0$.
        The conclusion is that co-planar detectors are not 
        sensitive to the circular polarization of the gravitational
        wave background. This is true of other types of detectors
        with planar geometries such as pulsar timing arrays and individual ground-based detectors such as LIGO \cite{LIGO:2009}. 
        
        To gain sensitivity to circular polarization we need to 
        introduce non-co-planar detector arms. Others 
        \cite{seto/taruya:2007} have considered
        using cross-correlations between ground-based detectors 
        like LIGO, VIRGO \cite{VIRGO:2012}, and KAGRA \cite{kagra},
        which have a suitable geometry. In the next subsection we 
        consider an extension to LISA in which we add a second 
        constellation of three satellites to break the co-planar
        geometry.

        \subsubsection{Two-constellations}
        \label{subsubsec:two_constellation}

        The extended LISA set-up is shown in the right 
        panel of Figure \ref{fig:interferometer_setup}. It consists 
        of two constellations of three equal-arm detectors. The two 
        constellations are separated by a rotation of $\pi$ radians 
        and a translation of $DL\hat{z}$. The $j^{\rm th}$ detector 
        on the $i^{\rm th}$ constellation is at position 
        $\mathbf{x}_{ij}$, and the unit vectors joining them are 
        given by: $u_i = (\mathbf{x}_{i2} - \mathbf{x}_{i1})/L, 
        \ v_i = (\mathbf{x}_{i3} - \mathbf{x}_{i1})/L, \ w_i = 
        (\mathbf{x}_{i3} - \mathbf{x}_{i2})/L$. We base this analysis
        on the designs proposed by Ref. \cite{smith/caldwell:2016} which 
        optimize the parameters $L$ and $D$ to achieve equal sensitivity 
        to intensity and polarization of the gravitational wave background. 
        Similar designs have also been considered by
        \cite{cornish:2001, crowder/cornish:2005, cornish/larson:2001}. 

        We use the signals defined in Equation \ref{eq:michelson}, but
        $\alpha, \beta$ are now written $\alpha_i, \beta_i$ where $i$ 
        refers to the constellation on which we are measuring the signal.
        The detector transfer functions are the same as the 
        single-constellation , but with extra indices referring to 
        the constellation we are considering 
        \cite{cornish:2001, smith/caldwell:2016}:
        \begin{equation}
        \begin{aligned}
        D^{\alpha_i}(f, \hat{u}_i \cdot \hat{n}) 
        &= \frac{1}{2}\exp(
        -2\pi i f \hat{n}\cdot \mathbf{x}_{i1}) \left[ \hat{u}_i 
        \otimes \hat{u}_i \ T_{\rm MI}(f, \hat{n} \cdot \hat{u}_i) 
        - \hat{v}_i \otimes \hat{v}_i \ T_{\rm MI}(f, \hat{n} \cdot 
        \hat{v}_i) \right], \\
        D^{\beta_i}(f, \hat{u}_i \cdot \hat{n}) 
        &= D^\alpha(f, \hat{u}_i \cdot \hat{n}) + \exp(-2\pi i f 
        \hat{n}\cdot \mathbf{x}_{i3}) \left[ \hat{u}_i \otimes 
        \hat{u}_i \ T_{\rm MI}(f, \hat{n} \cdot -\hat{u}_i) - 
        \hat{w}_i \otimes \hat{w}_i \ T_{\rm MI}(f, \hat{n} 
        \cdot - \hat{w}_i) \right]. \\
        \end{aligned}
        \end{equation}
        Following \cite{smith/caldwell:2016} we then combine Equations
        \ref{eq:michelson} to form estimators sensitive to just intensity
        or circular polarization:
        \begin{equation}
        \label{eq:IVestimators}
        \begin{aligned}
        \frac{1}{2} \delta_T(f - f\pr) \Rcal_I(f)I(f) &\equiv \langle \left[
        s^{\alpha_1}(f) + s^{\beta_1}(f) \right] \left[ s^{\alpha_2 *}
        (f\pr) + s^{\beta_2 *}(f\pr) \right] \rangle, \\
         \frac{1}{2} \delta_T(f - f\pr)\Rcal_V(f)V(f) &\equiv \langle 
        s^{\alpha_1}(f)s^{\beta_2 *}(f\pr) - s^{\beta_1}(f) 
        s^{\alpha_2 *}(f\pr) \rangle .
        \end{aligned}
        \end{equation}
        
         We will consider two experimental configurations of the two-constellation
         , introduced
         in Ref. \cite{smith/caldwell:2016}:  `LISA' with $L = 1\times10^9 \ 
         {\rm m}, \ D = 7,\ T = 10 \ {\rm years}$, and `BBO' with $L = 5 \times 
         10^7 \ {\rm m}, \ D = 2,\ T = 10 \ {\rm years}$. These designs are 
         optimized to achieve roughly equal sensitivity to $I$ and $V$.
        
          \subsection{Interferometer signal-to-noise}
          \label{subsec:sndesign}
          Under the assumption that the signals we are cross-correlating 
          have independent noise properties and are Gaussian-distributed, 
          and that the noise spectrum dominates over the signal, then
          the signal-to-noise in the interferometer is given by 
          \cite{smith/caldwell:2016, cornish:2001, romano/cornish:2016}:
          \begin{equation}
          \label{eq:interferometer_sn}
          \left( \frac{S}{N} \right)_{I,V}^2 = 2T\int_{0}^\infty df 
          \left(\frac{3H_0^2}{4\pi^2}\right)^2 
          \frac{|\Rcal_{\rm I, V}(f)\Omega_{\rm GW}^{I,V}(f)|^2}{f^6S^{I,V}_n(f)^2},
          \end{equation}
          where $S^{I,V}_n(f)$ is the power spectrum of the noise in the $I, 
          V$ signals, and $\Omega_{\rm GW}^{I,V}$ is the fractional energy 
          density of gravitational waves in intensity and circular polarization 
          today, defined in Equation \ref{eq:Odef}. To find the background 
          fractional energy density today we multiply the primordial spectrum
          by the appropriate transfer function
           \cite{barnaby/etal:2012, garcia-bellido/etal:2016, boyle/steinhardt:2008}:
          $  \Omega^{I,V}_{\rm GW}(f) = \frac{\Omega_{R,0}}{24}(\Pcal^L(f)\pm 
          \Pcal^R(f))$, where $\Omega_{R,0}$ is the fractional energy density
          in radiation today. \\
        
          Up to this point we have not discussed the noise, since it vanishes in
          the cross-correlations we consider. However it still 
          contributes to the variance of the estimators in Equations 
          \ref{eq:IVestimators}. There are three major sources of noise in 
          measurements of a particular optical path through an 
          interferometer: shot noise $S_{n,s}(f)$, accelerometer noise $S_{n,a}(f)$, and 
          the dominant laser phase noise, $S_{n,\phi}(f)$. As pointed out in 
          \S\ref{subsec:interferometer_response} the major motivation for 
          using equal-arm Michelson interferometers, as given in the first 
          two lines of Equations \ref{eq:michelson}, is the cancellation of 
          the laser phase noise. The shot and acceleration noises can be approximated
          by taking the fiducial LISA \cite{amaro-seoane/etal:2013} and
          BBO \cite{crowder/cornish:2005} values and 
          scaling them to an instrument with arm length $L$ observing at 
          frequency $f$ \cite{cornish:2001}. The final expressions for 
          $S_n^{I,V}(f)$ are derived by Ref. \cite{smith/caldwell:2016}:
          \begin{equation}
          \begin{aligned}
          S^I_n(f) &= \frac{121}{4} \left[  S_{n,s}(f) + 2S_{n,a}(f)\left(1+\cos^2
          \left(\frac{f}{f_*}\right)\right) \right]^2, \\
          S^V_n(f) &= \frac{96}{121}S^I_n(f),
          \end{aligned}
          \end{equation}
          where the values for $S_{n,a}(f)$ and $S_{s,a}(f)$ for LISA and BBO
          are given in Ref. \cite{smith/caldwell:2016}.
          As is the case for the CMB, our Galaxy contains sources of gravitational
          waves that may act as a confusion noise to a measurement of the GWB 
          \cite{farmer/phinney:2003, sathyaprakash/schutz:2009}. It 
          is expected that compact binary systems in our Galaxy will form a 
          gravitational wave foreground with an amplitude in intensity of
          $\Omega_{WD}\sim10^{-12}$ in the mHz regime. The shape of this spectrum 
          is quite complicated because different periods of a binary system's evolution
          dominate at different frequencies and have different frequency dependences \cite{farmer/phinney:2003}.  
          For the design of LISA we consider we expect the impact of
          such a foreground to be small compared to the acceleration noise 
          \cite{klein/etal:2016}. The BBO design we consider peaks at $\gtrsim 0.3$ 
          Hz, which is expected to be relatively free of such sources of noise 
          \cite{ferrari/etal:1999, ungalleri/vecchio:2001}. However, we are mainly
          interested in detecting chirality of the GWB, and this is more easily 
          distinguished from astrophysical foregrounds, and accordingly previous studies
          have not considered polarised foregrounds \cite{smith/caldwell:2016,
          seto:2007, garcia-bellido/etal:2016}. Therefore, we do not 
	      consider a contribution to the noise from astrophysical foregrounds in intensity 
	      or in polarisation, but it should be noted that we expect a small degradation 
	      in the achievable intensity sensitivity of the fiducial LISA design compared 
	      to our result, due to the confusion noise of astrophysical sources. 
           
 \subsection{Interferometer results}
 \label{subsec:interferometer_results}
 
  \begin{figure*}
        \centering
        	\includegraphics[]{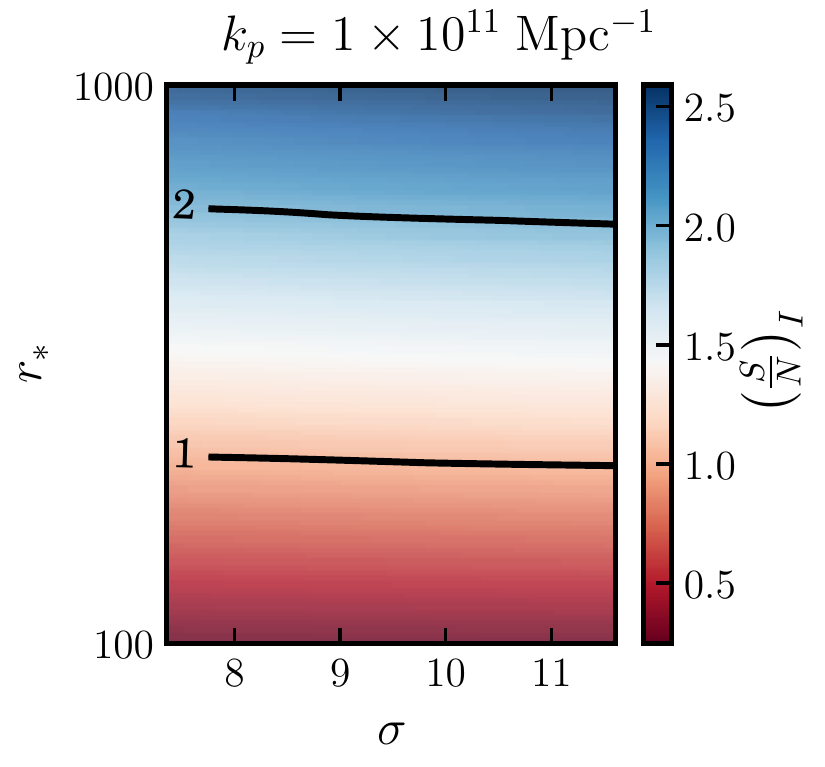}
    	\centering
    	\includegraphics[]{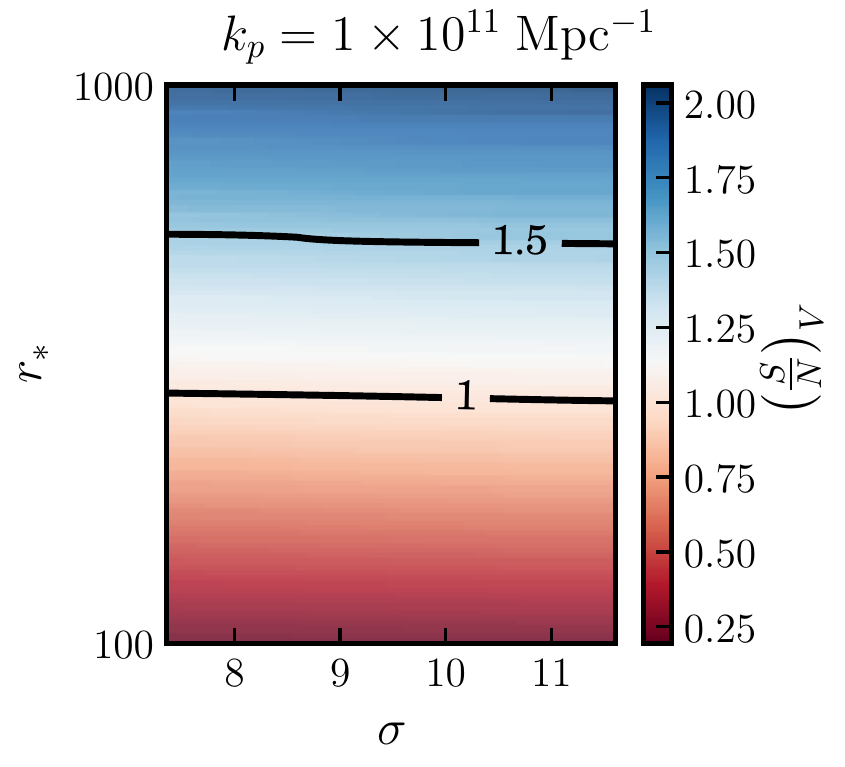}
        \caption{Signal-to-noise contours obtained using Equation 
                \ref{eq:interferometer_sn} for a LISA-like experiment described 
                in \S \ref{subsec:sndesign}. The primordial spectrum has $k_p = 1\times10^{11} 
                \perMpc $.} 
        \label{fig:LISA_contour}
  \end{figure*}
  
  \begin{figure*}
        \centering
        \includegraphics[]{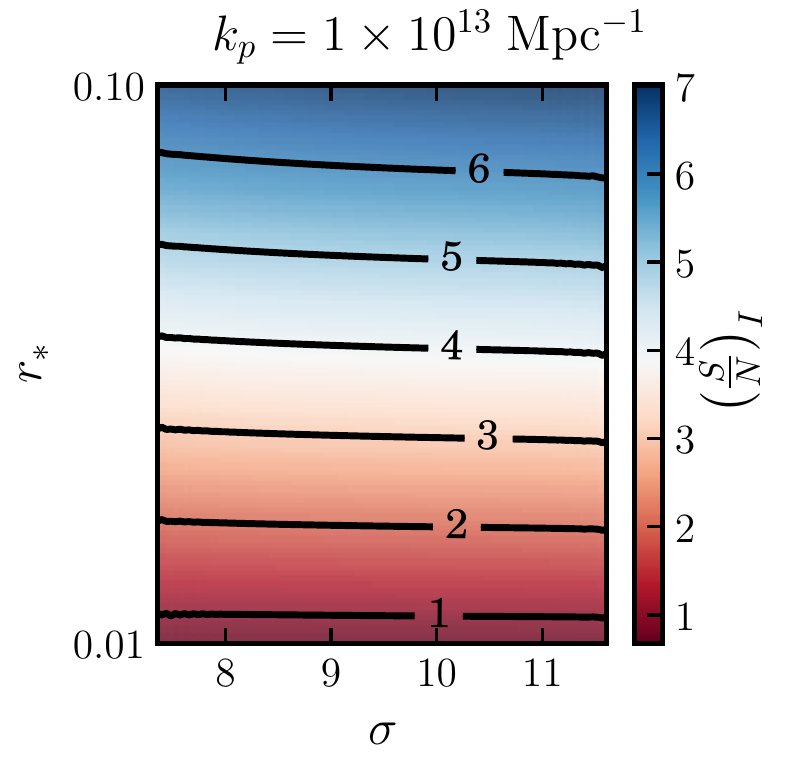}
        \includegraphics[]{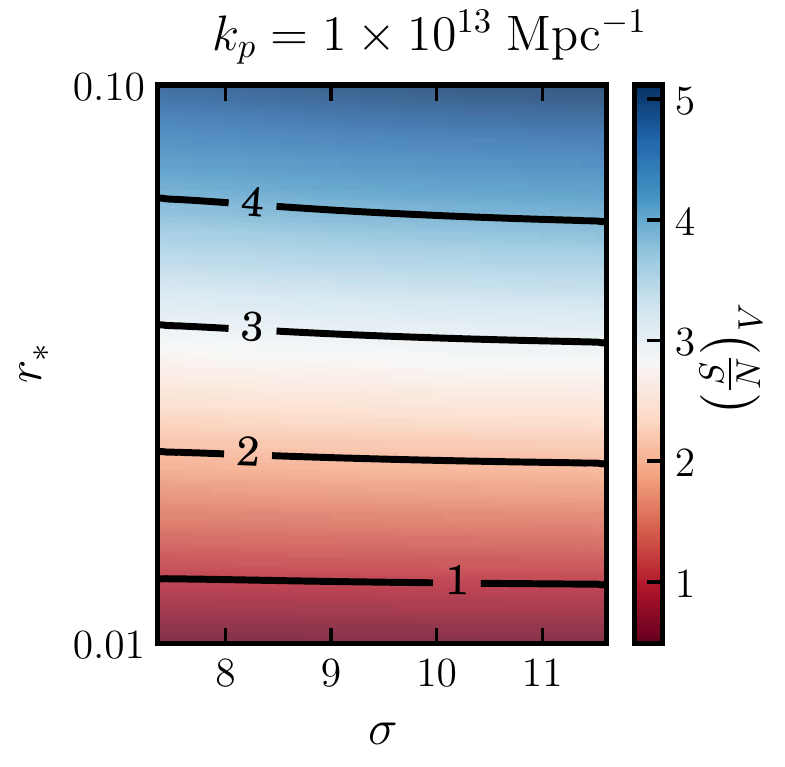}
        \caption{Signal-to-noise contours obtained using Equation 
                \ref{eq:interferometer_sn} for a BBO-like experiment described 
                in \S \ref{subsec:sndesign}. The primordial spectrum has $k_p = 1\times10^{13} 
                \perMpc $.}
        \label{fig:BBO_contour}
  \end{figure*}
          
  In Figure \ref{fig:LISA_contour} we plot signal-to-noise 
  contours for LISA assuming $k_p = 1 \times 10^{11} \perMpc$ and in Figure \ref{fig:BBO_contour}  we plot the 
  corresponding contours for BBO assuming $k_p = 1\times10^{13} 
  \perMpc $. We see that both the LISA and BBO 
  configurations may detect a polarized gravitational wave 
  background with signal-to-noise greater than one in a regime 
  unavailable to the CMB. In the case of LISA the signal-to-noise for 
  $k_p = 1\times10^{13} \perMpc$ is of order one. However, we see that 
  a BBO-like design far exceeds the sensitivity of LISA, probing a much 
  larger range of $r$ for the large $k_p$ values, inaccessible to CMB experiments. 
  A single 
  constellation design described in \ref{subsubsec:one_constellation}
  would achieve equivalent sensitivity in $I$ to LISA and BBO, but with no 
 $V$ sensitivity. Therefore, the fiducial LISA design would be sensitive
 to the inflationary model we consider here, since a positive detection of $I$
 at these scales with no corresponding detection on CMB scales would 
 require a strong scale dependence of the gravitational wave spectrum.
 
\section{Discussion}
\label{sec:discussion} 
 
 In this paper we have considered for the first time the detectability 
 of a new model for the production of gravitational waves proposed in Ref. 
 \cite{dimastrogiovanni/etal:2016}. Given the increasing effort to measure 
 the B-mode spectrum of the CMB, this is an important step in establishing 
 the origins of any detected primordial tensor perturbations. 
 This model has a unique tensor spectrum characterized by its 
 scale-dependence and chirality, both of which we use in order to find 
 observational markers that allow it to be distinguished from  the conventional 
 primordial gravitational waves produced by vacuum fluctuations. If a 
 detection of primordial gravitational waves is made, and the markers 
 we find to be detectable are absent, we may then rule out such a model. 
 In \S \ref{sec:cmb} we provided robust forecasts of the ability of 
 the LiteBIRD satellite mission to detect the TB and EB correlations 
 that result from the chiral tensor spectrum. We found that LiteBIRD would
 be able to detect the chirality for $r_* \gtrsim 0.03$, whilst $r_* 
 < 0.07$ is required by current observations. The addition of Stage 4 
 observations has little effect as such a survey would be limited to 
 $\ell > 30$, but the primordial chiral signal is contained almost entirely 
 within $2 < \ell < 30$. Further, we found that for cosmic-variance limited 
 observations the maximum achievable signal-to-noise for $r_*<0.07$ would 
 be $\sim 3$. From these studies we conclude that the ability of CMB two-point statistics to determine the presence of a chiral GWB is fairly limited. 
 
 However, in this study we have not fully leveraged the scale-dependence of the 
 axion-SU(2) model. Single-field slow-roll expects the tensor spectrum to
 have a tilt given by the self-consistency relation $n_T = -r/8$, and it would
 be possible to test departures from this using a combination of both
 CMB and interferometer constraints to provide a lever-arm 
 \cite{meerburg/etal:2015, lasky/etal:2016}. Such a study would be aided 
 by future groundbased observations such as Simons Observatory or S4. 
 In this study we found that for a peak wavenumber in the range $k_p \sim 7 
 \times 10^{-5} - 5 \times 10^{-3} \perMpc$  the primordial BB spectrum is 
 detectable by LiteBIRD with $(S/N)_{BB} \gtrsim 1$ for $r_* \gtrsim 10^{-3}$. 
 However, the projected sensitivity on $n_T$ for LiteBIRD alone is $\sim 0.04$, 
 which is not sufficient to test deviations from 
 the self-consistency relation, without external constraints.
 
 Another characteristic of the axion-SU(2) model of Ref. 
 \cite{dimastrogiovanni/etal:2016} is its intrinsic non-Gaussianity. 
 Some studies have recently shown that higher order statistics of B-modes, 
 such as the BBB bispectrum, may yield a $>2\sigma$ significance for the axion-U(1) model \cite{shiraishi/etal:2016, namba/etal:2016}. An analysis of 
 the CMB non-Gaussianity for the axion-SU(2) model is therefore in order \cite{agrawal/etal:2017}.
 \\

 In \S\ref{sec:interferometers} we showed that interferometers may provide
 a complementary probe to the CMB at much smaller scales $\sim 10^{12} 
 \perMpc$, even for the relatively flat spectra required by the attractor 
 behaviour of the background axion field coupled to the SU(2) gauge field. 
 This takes advantage of the scale-dependence of the axion-SU(2)
 model, which allows the spectrum to have a large excursion at some 
 scale $k_p$, e.g. as shown in Figure \ref{fig:comparison}, making the cosmological GWB
 of the axion-SU(2) model a viable target for interferometers with current 
 sensitivities. We went on to consider two designs of an advanced stage 
 LISA-like mission proposed by Ref. \cite{smith/caldwell:2016} which are 
 sensitive to both the intensity and circular polarization of the GWB. Whilst  
 interferometers are not in general sensitive to the same parameter space of 
 the model as CMB probes, 
 we found that for spectra with a very large values of $k_p$ and 
 $\sigma$, that would be undetectable on CMB scales, such experiments could 
 make significant detections, and therefore complement CMB constraints.

\begin{acknowledgments}
BT would like to acknowledge the support of the University of Oxford-Kavli IPMU Fellowship
and an STFC studentship. MS was supported in part by a Grant-in-Aid for JSPS Research under Grant No. 27-10917 and JSPS Grant-in-Aid for Research Activity Start-up Grant Number 17H07319.  The work of TF
is partially supported by the JSPS Overseas Research Fellowships, Grant No.~27-154. Numerical computations were in part carried out on Cray XC30 at Center for Computational Astrophysics, National Astronomical Observatory of Japan. We were supported in part by the World Premier International Research Center Initiative (WPI Initiative), MEXT, Japan. TF would like to thank Kavli IPMU for warm hospitality during his stay. This work was supported in part by JSPS KAKENHI Grant Number JP15H05896. MH and NK acknowledge support from MEXT KAKENHI Grant Number JP15H05891.
\end{acknowledgments}
 
\bibliography{superbib}
\appendix

\section{Derivation of the template for GW power spectrum}
\label{app:derivation}

\begin{figure*}
        \centering
        \includegraphics[width=0.49\textwidth]{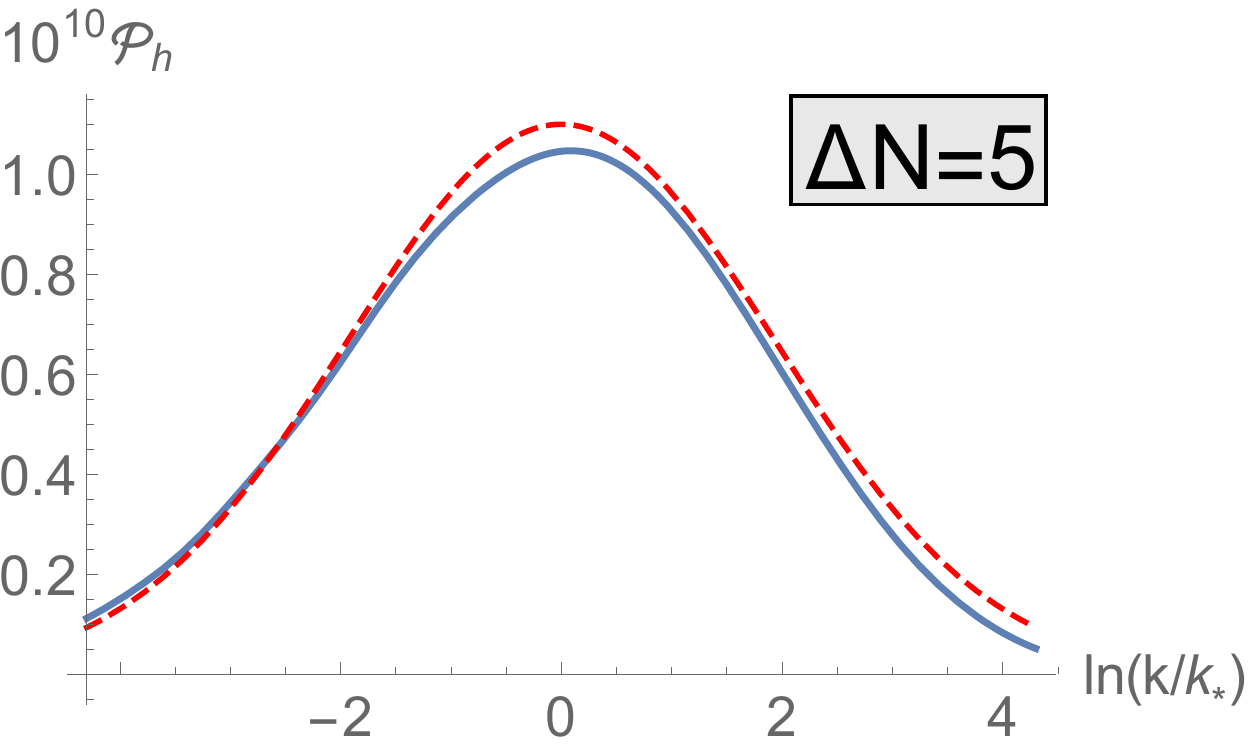}
        \includegraphics[width=0.49\textwidth]{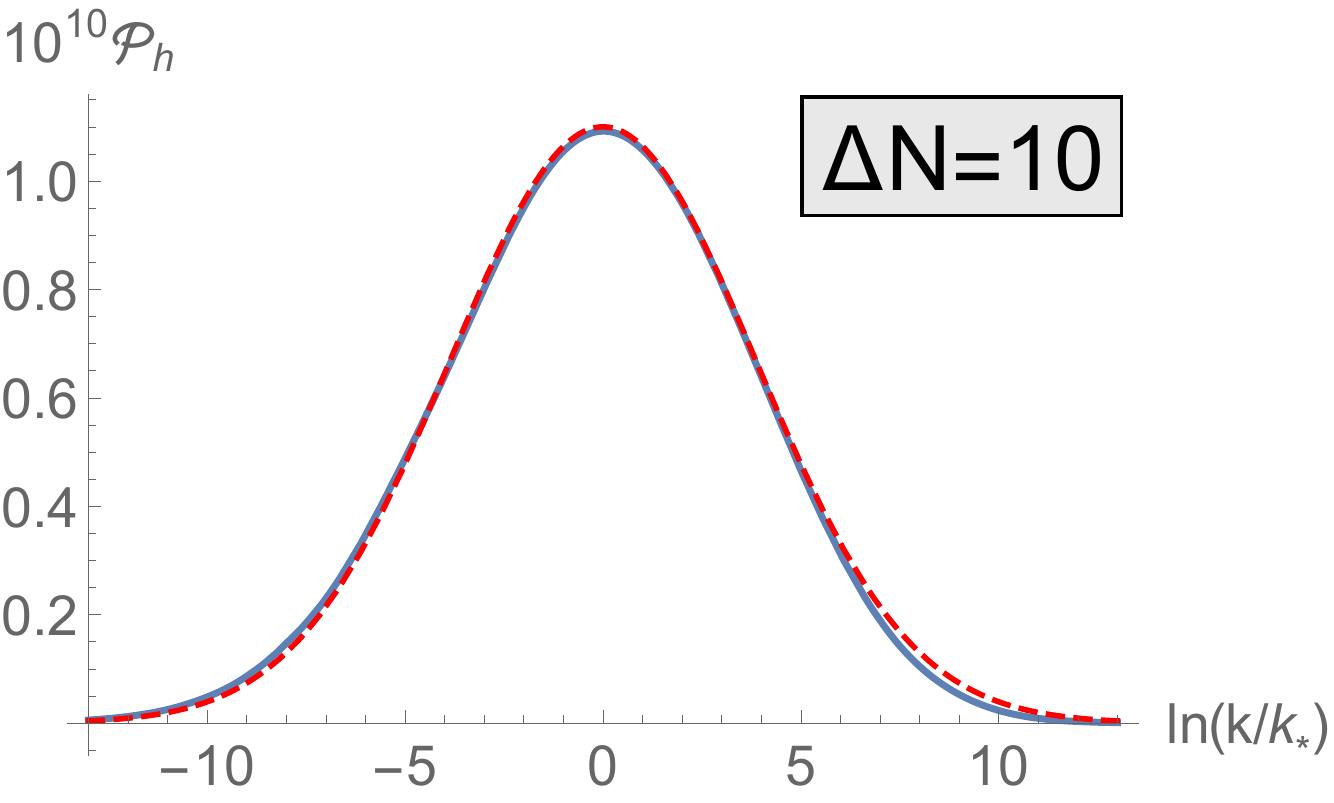}
        \caption{The comparison between the full numerical result of $\Pcal_h^{\rm Sourced}(k)$ (blue solid line) and the template eq.~\eqref{eq:fitting_formula} with eq.~\eqref{r* expression} and $\sigma^2 = 0.15 \Delta N^2$ (red dashed line). In the left (right) panel, $\Delta N = 5 (10), m_* =4, \epsilon_{B*}\approx 9\times 10^{-4}$ and the peak amplitude reaches the tensor-to-scalar ratio, $r_*=0.05$.
                The Hubble parameter is set as $H_{\inf}=8\times 10^{11}\ {\rm GeV}$ which corresponds to $r=10^{-5}$ without the sourced GW. In the case of $\Delta N=5$, the derived formula slightly underestimate the peak amplitude and the width,
                while the fit is excellent for $\Delta N\gtrsim 10$.}  
        \label{fig:template_check}
\end{figure*}

In Ref.\cite{dimastrogiovanni/etal:2016}, it has been shown that the power spectrum of the sourced GW is given by 
\begin{equation}
\mathcal{P}_h^{\rm L,Sourced}(k) = \frac{\epsilon_B H_{\inf}^2}{\pi^2 M_{\rm Pl}^2}\mathcal{F}^2(m_Q),
\label{DFF Ph}
\end{equation}
where $H_{\inf}$ is the inflationary Hubble scale, $\epsilon_B \equiv g^2Q^4/(M_{\rm Pl}^2 H_{\inf}^2)$ roughly indicates the energy fraction of the SU(2) gauge field. $\mathcal{F}(m_Q)$ is a monotonically increasing  function for $3\le m_Q \le 7$ which
is well approximated by 
\begin{equation}
\mathcal{F}(m_Q) \simeq \exp\left[2.4308m_Q-0.0218m_Q^2-0.0064m^3_Q-0.86\right],
\quad (3\le m_Q \le 7),
\label{F fitting}
\end{equation}
where the value of a dynamical parameter $m_Q(t)\equiv g Q(t)/H_{\inf}$ around the horizon crossing $k\sim aH_{\inf}$ is substituted. 
Solving the background equations of motion for $\chi(t)$ and $Q(t)$ with the slow-roll approximation, one can show
\begin{equation}
m_Q(t) =m_* \sin^{1/3}\left[\chi(t)/f \right].
\label{mQt}
\end{equation}
where $m_* \equiv \left(g^2 \mu^4/3\lambda H^4_{\inf}\right)^{1/3}$ is the maximum value of $m_Q(t)$.
From the definition of $m_Q$ and $\epsilon_B$, the value of $\epsilon_B$ at $m_Q=m_*$ is $\epsilon_{B*} \equiv H^2 m_*^4/g^2 M_{\rm Pl}^2 $.
Therefore the tensor-to-scalar ratio $r$ on the peak scale $k_p$ of the sourced GW power spectrum is%
\begin{equation}
r_* = \frac{\mathcal{P}_h^{\rm L,Sourced}}{\Pcal_\zeta}(k_p) = \frac{\epsilon_{B*} H_{\inf}^2}{\pi^2 M_{\rm Pl}^2 \Pcal_\zeta}\mathcal{F}^2(m_*).
\label{r* expression}
\end{equation}
Next, we consider the width of the GW spectrum. Around the peak of $m_Q(t)$ at $t=t_*$, or $\chi(t=t_*)=\pi f/2$, $\chi(t)$ is expanded as
\begin{align}
\chi(t)\simeq \frac{\pi}{2}f +\dot{\chi}_* (t-t_*) 
\simeq f\left[ \frac{\pi}{2} + \frac{2\xi_*}{\lambda} H_{\inf}(t-t_*)\right],
\label{chi expansion}
\end{align}
where $\dot{\chi}_* \equiv \dot{\chi}(t=t_*)$, $\xi_*\equiv \lambda \dot{\chi}_*/(2fH_{\inf})$ and one can show $\xi_*\simeq m_*+m_*^{-1}$ in the slow-roll regime.
Then we obtain the approximated equation for $m_Q(t)$ which is valid around the peak value ,
\begin{equation}
m_Q(t)\simeq m_* \left[1-\frac{1}{6} \left( \frac{H_{\inf}(t-t_*)}{\Delta N}\right)^2\right],
\qquad (t\sim t_*),
\label{mQ approx}
\end{equation}
where we define $\Delta N \equiv \lambda/2\xi_{*}$. 
Substituting it into eq.~\eqref{DFF Ph} and using $H_{\inf}(t-t_*)=\ln(k/k_p)$, we obtain the leading order result as 
\begin{equation}
\mathcal{P}_h^{\rm L,Sourced}(k) \simeq \frac{\epsilon_{B*} H_{\inf}^2}{\pi^2 M_{\rm Pl}^2}\mathcal{F}^2(m_*) \times \exp\left[-\mathcal{G}(m_*)\frac{\ln^2(k/k_p)}{\Delta N^2} \right],
\end{equation}
with $\mathcal{G}(m_*)\approx 0.666+ 0.81m_*-0.0145m_*^2-0.0064m_*^3$.
Note that the contribution from $\epsilon_B(t)\propto m_Q^4(t)$ in the prefactor should not be missed. Comparing it with the template eq.~\eqref{eq:fitting_formula}, one finds 
\begin{equation}
\sigma^2=\frac{\Delta N^2}{2\mathcal{G}(m_*)}.
\end{equation}
The validity of the derived expression for $\mathcal{P}_h^{\rm L,Sourced}(k) $ is checked by the comparison with the full numerical result. 
Once $r_*, m_*, \epsilon_{B*}$ and $\Delta N$ are fixed, all the model parameters
$g, \lambda, \mu$ and $f$ are determined. Then we  can numerically solve the background equation of $\chi(t)$ and $Q(t)$ as well as the equations for the perturbations $t_L(k,t)$ and $h_L(k,t)$ to obtain the power spectrum of the sourced GW. 
In Figure.~\ref{fig:template_check}, we compare the derived expression with the full numerical result. 
It should be noted that eq.~\eqref{DFF Ph} and our derivation rely on the slow-roll approximation. The approximation is less accurate for a small $\Delta N$, because $\Delta N$ characterizes the time scale of $\chi(t)$ rolling down its potential.
In Figure.~\ref{fig:template_check}, one can find a small deviation in the case of $\Delta N=5$, while the excellent agreement is seen for $\Delta N=10$.
% 
%///////////////////////////////////////////////////////////////////////////////////%
%///////////////////////////////////////////////////////////////////////////////////%

\begin{figure}
\centering
\includegraphics[]{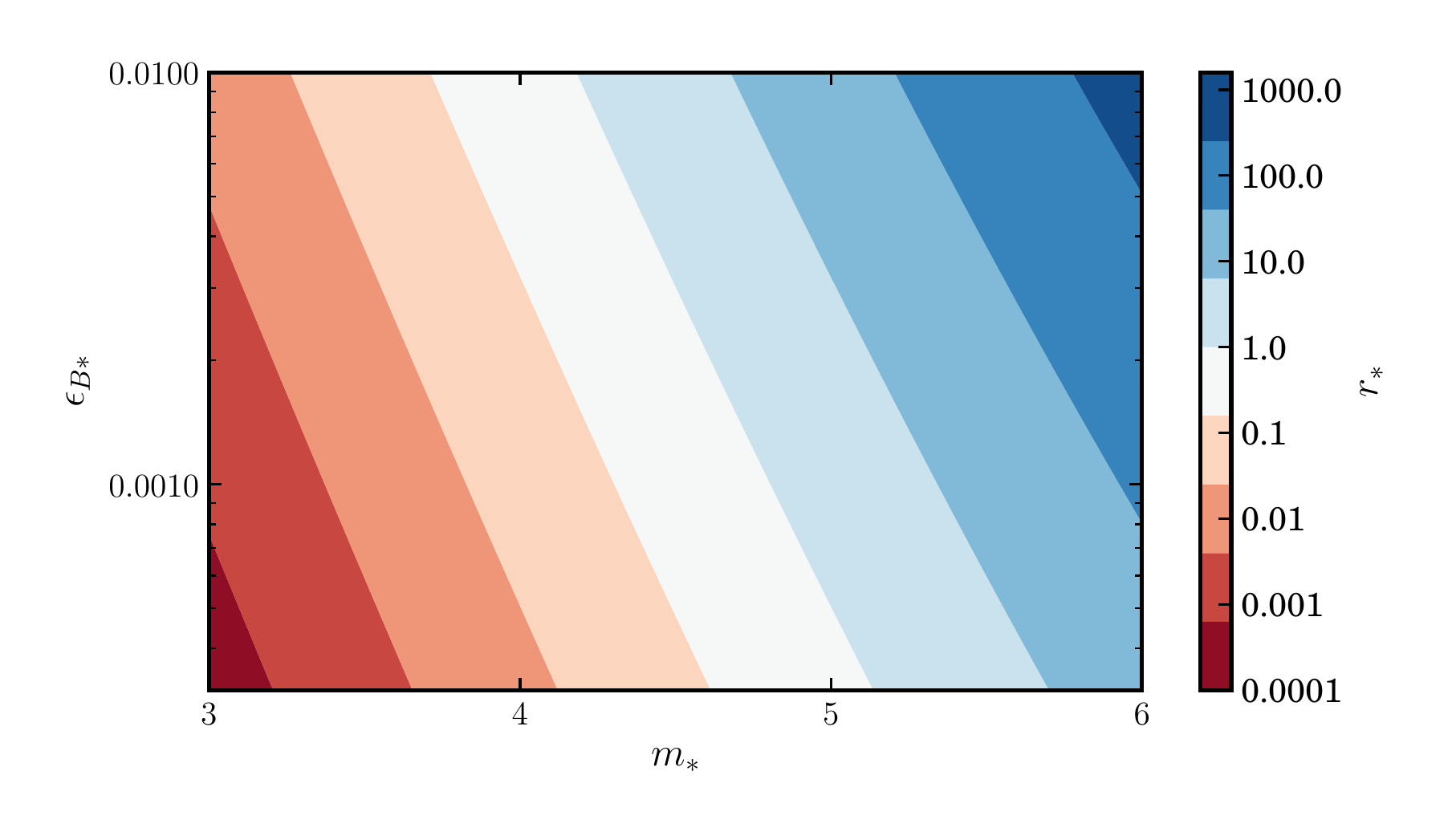}
\caption{Peak tensor-to-scalar ratio $r_*$ as a function of $\epsilon_{B*}$ and 
        $m_*$ for $k_p = 0.005 \perMpc$.}
\label{fig:CVebm}
\end{figure}

Finally, we discuss how long it takes  $\chi$ to get to $\chi_*$,
given that the initial value of $\chi$ is negligibly small compared to $f \pi/2$.
Assuming $\chi(t\approx 0)\ll f$ and using eq.~\eqref{chi expansion}, one finds
\begin{equation}
H t_* \sim\frac{\pi}{2}\Delta N.
\end{equation}
However, it is definitely underestimated, because $\dot{\chi}(t\ll t_*)$
must be smaller than $\dot{\chi}(t=t_*)$ which is the maximum value.
In fact, a full numerical calculation shows that the coefficient is somewhat larger,
\begin{equation}
H t_* \approx 1.8 \Delta N.
\end{equation}
One may wonder if $\chi(t)$ can stay on the top of its potential hill for a longer time if its initial value is small enough. However, since $\chi$ is coupled to the SU(2) gauge fields and the system quickly goes to the attractor behavior, the time scale of the motion of $\chi$ is almost solely determined by $\Delta N$. 
It indicates that the peak scale $k_p$ should be smaller than $k_i \exp[1.8\Delta N]$. Here $k_i$ is the wave number of the mode exiting the horizon at the initial time, and it is smaller or roughly equals to the largest CMB scale.
Therefore we obtain the following constraint on $\Delta N$,
\begin{equation}
\Delta N \gtrsim \frac{1}{1.8}\ln\left(\frac{k_{p}}{k_{\rm CMB}}\right).
\label{dN constraint}
\end{equation}

\vspace{1cm}

\section{Calculation of the covariance matrix, $\xi$}
\label{app:covariance_matrix}
For a given beam, $b_\ell$, and a white noise level, 
$w^{-1}_{X_1X_2}$, the expected variance of the 
multipoles of an observed sky is given by:
\begin{equation}
\langle (a_{\ell m}^{X_1})^* a_{\ell^\prime 
	m^\prime}^{X_2} \rangle = \left(|b_\ell|^2 C_\ell
^{X_1X_2}+ w^{-1}_{X_1X_2} \right)\delta_{\ell 
	\ell^\prime}\delta_{m m^\prime}.
\end{equation}
An unbiased estimator of the angular power spectrum 
is then:
\begin{equation}
\hat C_\ell^{X_1X_2} = |b_\ell|^2 \left( \sum_{m=
	-\ell}^\ell \frac{(a_{\ell m}^{X_1})^* a_{\ell^\prime 
		m^\prime}^{X_2}}{2\ell+1} - w^{-1}_{X_1X_2} \right)
\end{equation}
By considering the expectation $\langle (\hat{C}
^{X_1X_2}_\ell - C^{X_1X_2}_\ell)(\hat{C}^{X_3X_4}_\ell 
- C^{X_3X_4}_\ell)  \rangle$ it can then be shown that 
the covariance is given by 
\cite{glusevic/kamionkowski:2010}:
\begin{equation}
\xi^{X_1X_2X_3X_4} = \frac{1}{(2\ell+1)f_{\rm sky}}(\tilde C_\ell
^{X_1X_3}\tilde C_\ell^{X_2X_4} + \tilde C_\ell^{
	X_1X_4}\tilde C_\ell^{X_2X_3}).
\end{equation}
where $\tilde C_\ell^{X_1X_2} = C_\ell^{X_1X_2} + 
|b_\ell|^{-2}w^{-1}_{X_1X_2}$. 

\vspace{1cm}

\section{CMB noise spectrum}
\label{app:aggregate_noise}

  For a given set of experimental parameters such as channel 
  frequencies, FWHM and sensitivity in polarization and temperature 
  per channel we want to find the aggregate noise in the CMB spectra. 
  We follow the treatment of Ref. \cite{shiraishi/etal:2016}, which 
  itself closely follows Ref. \cite{oyama/etal:2016}.

  There are multiple sources of noise in the final spectrum: 
  instrumental noise in the CMB channels, residual foreground noise 
  from incomplete cleaning, and additional systematic noise 
  introduced from the templates used in cleaning the CMB channels.

  The noise in the final CMB spectrum is:
  \begin{equation}
  N_\ell^{\rm BB} = \left[ \sum_i \frac{1}{n_\ell(\nu_i) + \left[ 
  C_\ell^{\rm S}(\nu_i) + C_\ell^{\rm D}(\nu_i)\right] \sigma_{\rm 
  RF} + n_\ell^{\rm RF}(\nu_i)}\right]^{-1}
  \end{equation}
  where the index $i$ runs over channels used in CMB analysis, RF 
  refers to residual foregrounds, $n_\ell(\nu)$ is the noise 
  spectrum in the channels used for CMB analysis, $\left[ C_\ell
  ^{\rm S}(\nu_i) + C_\ell^{\rm D}(\nu_i)\right] \sigma_{\rm RF}$ 
  is the residual foreground level in dust and synchrotron 
  rescaled to the frequencies used in CMB analysis, and $n_\ell^{
  \rm RF}(\nu_i)$ is the instrumental uncertainty in the process of 
  foreground removal. 

  The simplest of the above terms is the noise in the CMB channels:
  \[
  n_\ell(\nu) = \sigma_P^2(\nu) \exp \left[ \frac{\ell(\ell+1)
  \left(\frac{\pi}{10800}\theta_{\rm FWHM}(\nu)\right)^2} {8 \ln (2)} \right],
  \]
	where $\theta_{\rm FWHM}(\nu)$ is the FWHM of the channel $\nu$ 
	in arcminutes. 
  The instrumental uncertainties in the process of foreground 
  removal are given by Ref. \cite{oyama/etal:2016}:
  \[
  n_\ell^{\rm RF} = \frac{4}{N_{\rm chan}(N_{\rm chan}-1)}\left[ 
  \sum_j \frac{1}{n_\ell(\nu_j)} \right]^{-1} \left[ \left( 
  \frac{\nu}{\nu_{S, ref}} \right)^{2\alpha_S} + \left( 
  \frac{\nu}{\nu_{D, ref}} \right)^{2\alpha_D}  \right] ,
  \]
  where $N_{\rm chan}$ is the number of channels used in 
  foreground cleaning (in this case $N_{\rm chan}=10$), and $\nu_{S, ref}, \nu_{D,ref}$ are the 
  highest and lowest frequency channel used in the removal (in 
  this case $\nu_{S, ref}=30 \ {\rm GHz},\ \nu_{D,ref}=94 \ {\rm GHz}$).
  The foreground spectra are:
  \[
  C_\ell ^S (\nu ) = A_S \left( \frac{\nu}{\nu_{S, 0}} 
  \right)^{2 \alpha_S}\left( \frac{\ell}{\ell_{S,0}} \right)^{
  \beta_S}  
  \]

\[
C_\ell^D(\nu ) = p^2 A_D \left( \frac{\nu}{\nu_{D, 0}} \right)^{2 
\alpha_D}\left( 
\frac{\ell}{\ell_{D,0}} \right)^{\beta_D}
 \left[ \frac{ e^{ \frac{h\nu_{D,0}}{k_BT} } -1 } { e^{\frac{h\nu}{k_BT}}-1 }
\right].
\]
These are converted into a Gaussian addition to the noise by the factor 
$\sigma^{\rm RF}$ such 
that a 2\% residual level corresponds to $\sigma^{\rm RF} = 4\times10^{-4}$. 

The spectral parameters of the foreground s are summarized in Table 
\ref{tab:spectral_params}.
They are taken from Ref. \cite{oyama/etal:2016}, and are consistent with the 2015 
Planck data.

\begin{table*}%[b]
    \centering{

      \begin{tabular}{cc}
      Parameter   & Value                \\
      \hline
      $A_S$       & $4.7 \times 10^{-5} \ {\rm \mu K}^2$         \\
      $\alpha_S$  & -3                   \\
      $\beta_S$   & -2.6                 \\
      $\nu_{S,0}$ & 30 GHz               \\
      $\ell_{S,0}$& 350                  \\
      \hline
      $A_D$       & $1 \ {\rm \mu K}^2$  \\
      $\alpha_D$  & 2.2                  \\
      $\beta_D$   & -2.5                 \\
      $\nu_{D,0}$ & 94 GHz               \\     
      $\ell_{D,0}$& 10                   \\
      \hline
      $T$         & 18 K                 \\
      $p$         & 0.15                 \\
      \hline
      \end{tabular}}
      \caption{Spectral parameters used in noise model taken from Ref.
\cite{shiraishi/etal:2016}.}
    \label{tab:spectral_params}
    \end{table*}
     
   \begin{table*}%[b]
    \centering{
      \begin{tabular}{ccc}
      \hline
      Channel (GHz) & $\theta_{\rm FWHM}$ (amin) & $\sigma_{\rm P}(\nu) \ \left[{\rm \mu K 
amin}\right]$ \\
      \hline
      40.0  & 69.0 & 36.8 \\
      50.0  & 56.0 & 23.6 \\
      60.0  & 48.0 & 19.5 \\
      68.0  & 43.0 & 15.9 \\
      78.0  & 39.0 & 13.3 \\
      89.0  & 35.0 & 11.5 \\
      100.0 & 29.0 & 9.0 \\
      119.0 & 25.0 & 7.5 \\
      140.0 & 23.0 & 5.8 \\
      166.0 & 21.0 & 6.3 \\
      195.0 & 20.0 & 5.7 \\
      235.0 & 19.0 & 7.5 \\
      280.0 & 24.0 & 13.0 \\
      337.0 & 20.0 & 19.1 \\
      402.0 & 17.0 & 36.9 \\
      \hline
      \end{tabular}}
      \caption{Summary of the ${\rm LiteBIRD}$ specifications ($f_{\rm sky} 
= 0.5$).}
    \label{tab:lbird_specs}
    \end{table*}
    \FloatBarrier
    
\section{Frequency dependence of CMB sensitivity}
\label{app:cmb_freqdep}

When we calculate the CMB angular power spectrum we are decomposing the signal
into multipoles corresponding to certain angular distance on the sky. Each 
multipole has contributions from all frequencies of the GWB, determined by an 
integral of transfer functions:
\[
C_\ell^{YY^\prime} = 4 \pi \int \frac{dk}{k} \left[ \mathcal{P}_h^{\rm L}(k)- 
\mathcal{P}_h^{\rm R}(k) \right]\Delta^h_{Y,\ell}(k)\Delta^h_{Y^\prime , 
\ell}(k).
\]
This makes a direct link between multipole and frequency ambiguous.
Since the transfer functions are sharply peaked at $k_\ell = \ell / \eta_0$ with $\eta_0$ denoting the comoving distance to the last scattering surface.
We make the approximation:
\begin{equation}
\begin{aligned}
C_\ell^{YY^\prime}(k_\ell) &= \left[4 \pi \int \frac{dk}{k} \left[ 
\mathcal{P}_h^{\rm L}(k, r_*=1)- 
\mathcal{P}_h^{\rm R}(k, r_*=1) \right]\Delta^h_{Y,\ell}(k)\Delta^h_{Y^\prime , 
\ell}(k)\right]
(\mathcal{P}_h^{\rm L}(k_\ell, r_*)- \mathcal{P}_h^{\rm R}(k_\ell, r_*))  \\ 
&= C_\ell^{YY^\prime}(r_*=1) (\mathcal{P}_h^{\rm L}(k_\ell, r_*)- 
\mathcal{P}_h^{\rm R}(k_\ell, r_*))
\end{aligned}
\end{equation}

To calculate the sensitivity to a circular background we calculate the 
signal-to-noise of the TB spectrum, ignoring the small contribution from 
EB for simplicity. The signal-to-noise is therefore:
\[
 (S/N)^2_{\rm TB, \ \ell} = (2 \ell + 1)f_{\rm sky} 
 \frac{(C^{TB}_\ell)^2}{\hat{C}^{TT}_\ell\hat{C}^{BB}_\ell},
\]
where over-hat indicates the observed spectrum, including foreground residuals, 
instrument noise, and lensing. Our assumption that the transfer function is 
strongly peaked at $k_\ell$ now allows us to write this as a function of 
$k_\ell$ instead of just $\ell$:
\[
 (S/N)^2_{\rm TB}(k_\ell) = (2 \ell + 1)f_{\rm sky} 
 \frac{(C^{TB}(k_\ell))^2}{\hat{C}^{TT}_\ell\hat{C}^{BB}_\ell}.
\]
Note that we still calculate the observed spectrum fully. We then ask the 
question: what is the required $P_h^{\rm L}(k_\ell)$ (take $P_h^R = 0$) to 
achieve a signal-to-noise of one in the channel $k_\ell$? This will be the 
minimum GWB detectable with a signal-to-noise of one. So:
\[
(\mathcal{P}_h^{\rm L}(k_\ell, r_*)- \mathcal{P}_h^{\rm R}(k_\ell, r_*))^{\rm 
min}  = 
\sqrt{ \frac{ \hat{C}^{TT}_\ell\hat{C}^{BB}_\ell }{(2 \ell + 1)f_{\rm sky}}} 
[C_\ell^{TB}(r_*=1)]^{-1} .
\]
This quantity tells us about the tensor spectrum at recombination, however in 
order to compare with interferometers which are sensitive to the current GWB, 
we have to evolve this forward in time. The tensor spectrum transfer function 
for CMB scales is \cite{smith/caldwell:2016, watanabe/komatsu:2006}:
\begin{equation}
\label{eq:cmb_transfers}
\Omega^{\rm min}_Vh^2 = 1875 (\Pcal_h^{\rm L}(k_\ell)-\Pcal_h^{\rm R}(k_\ell))^{\rm min}
\left(\frac{3j_2(k_\ell 
\eta_0)}{k_\ell \eta_0}\frac{k_\ell}{k_*} \right)^2
\end{equation}

\end{document}